\documentclass[journal,twoside,web]{ieeecolor}
\usepackage{jsen}
\usepackage{amsmath,amsfonts,mathtools}
\usepackage{algorithmic}
\usepackage{algorithm}
\usepackage{array}
\usepackage{textcomp}
\usepackage{stfloats}
\usepackage[hyphens]{url}
\usepackage{verbatim}
\usepackage{graphicx}
\usepackage{tabularx, booktabs, multirow}
\usepackage[noadjust]{cite}
\usepackage{balance}
\usepackage[skip=3pt]{subcaption}
\usepackage{wrapfig}
\usepackage[hidelinks]{hyperref}

\newcommand{\norm}[1]{\left\lVert #1 \right\rVert}
\newcommand{\Norm}[1]{\lVert #1 \rVert}

\newcommand{\set}[2]{\left\{ \, #1 \mid #2  \, \right\}}
\newcommand{\Set}[2]{\{ \, #1 \mid #2  \, \}}
\newcommand{\abs}[1]{\left\lvert #1 \right\rvert}

\DeclareMathOperator*{\argmin}{arg\,min}
\DeclareMathOperator*{\argmax}{arg\,max}
\DeclareMathOperator{\trace}{Tr}

\DeclareMathOperator{\diag}{diag}

\newtheorem{problem}{Problem}

\newtheorem{theorem}{Theorem}
\newtheorem{lemma}{Lemma}

\def\BibTeX{{\rm B\kern-.05em{\sc i\kern-.025em b}\kern-.08em
    T\kern-.1667em\lower.7ex\hbox{E}\kern-.125emX}}
\definecolor{abstractbg}{rgb}{0.89804,0.94510,0.83137}
\definecolor{nu_green}{rgb}{0, .6, 0.4}

\begin{document}
\begin{figure*}
This paper was published in \href{https://doi.org/10.1109/JSEN.2025.3537702}{\journalname, Vol.~25, No.~6, pp.~10030--10045, 2025, doi: 10.1109/JSEN.2025.3537702}.\\[10pt]
©2025 IEEE.  Personal use of this material is permitted.  Permission from IEEE must be obtained for all other uses, in any current or future media, including reprinting/republishing this material for advertising or promotional purposes, creating new collective works, for resale or redistribution to servers or lists, or reuse of any copyrighted component of this work in other works.
\end{figure*}
\title{Fast Data-driven Greedy Sensor Selection \\ for Ridge Regression}

\author{Yasuo Sasaki, Keigo Yamada, Takayuki Nagata, Yuji Saito, and Taku Nonomura
\thanks{This work was supported by Japan Science and Technology Agency (JST) Moonshot R\&D Program under Grant JPMJMS2287, JST CREST under Grant JPMJCR1763, and Japan Society of the Promotion of Science KAKENHI under Grant 21J14180 and 23K13348.}%
\thanks{Y. Sasaki, T. Nagata, and T. Nonomura are with the Department of Aerospace Engineering, Nagoya University, Nagoya, 4648603, Japan (e-mail: sasaki.yasuo.g8@f.mail.nagoya-u.ac.jp; takayuki.nagata@mae.nagoya-u.ac.jp; nonomura@nagoya-u.jp).
K. Yamada is with the Department of Aerospace Engineering, Tohoku University, Sendai, 9808579, Japan (e-mail: keigo.yamada.t5@dc.tohoku.ac.jp).
Y. Saito is with the Frontier Research Institute for Interdisciplinary
Sciences, Tohoku University, Sendai, 9808578, Japan (e-mail: yuji.saito@tohoku.ac.jp)}
}

\IEEEtitleabstractindextext{%
\fcolorbox{abstractbg}{abstractbg}{%
\begin{minipage}{\textwidth}%
\begin{wrapfigure}[16]{r}{3in}%
\vspace{-10pt}%
\includegraphics[width=3in]{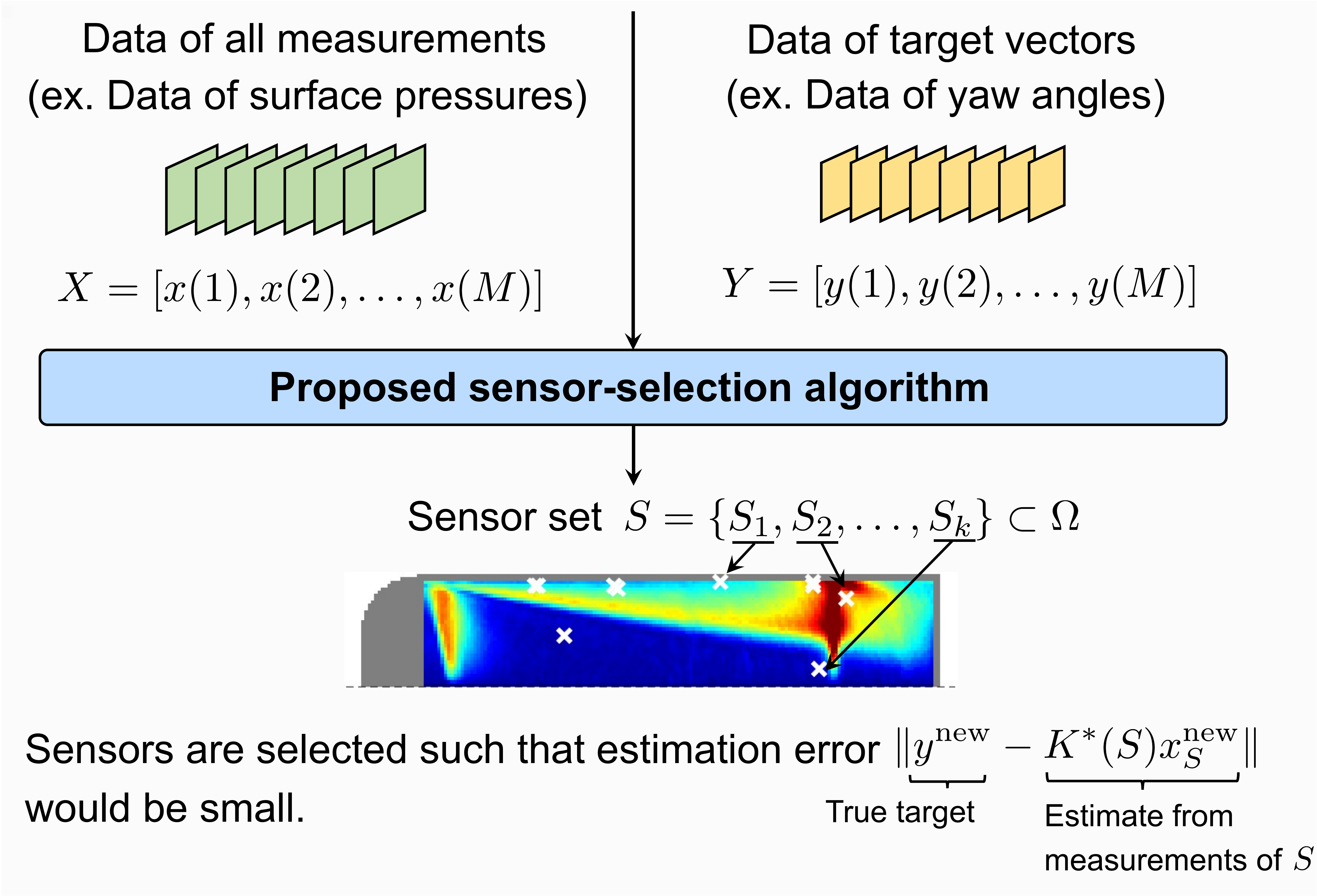}%
\end{wrapfigure}%
\begin{abstract}
We propose a data-driven sensor-selection algorithm for accurate estimation of the target variables from the selected measurements.
The target variables are assumed to be estimated by a ridge-regression estimator which is trained based on the data. 
The proposed algorithm greedily selects sensors for minimizing the cost function of the estimator.
Sensor selection which prevents overfitting of the resulting estimator can be realized by setting a positive regularization parameter.
The greedy solution is computed in quite a short time by using some recurrent relations that we derive.
The effectiveness of the proposed algorithm is verified for artificial datasets which are generated from linear systems and a real-wold dataset which are aimed for selection of pressure-sensor locations for estimating yaw angle of a ground vehicle.
The demonstration for the datasets reveal that the proposed algorithm computes a sensor set resulting in more accurate estimation than existing data-drive selection algorithms in some conditions.
Furthermore, it is confirmed that setting a positive regularization parameter in the proposed algorithm leads to accurate estimation when overfitting is problematic.
\end{abstract}

\end{minipage}}}

\maketitle

\section{Introduction}
\IEEEPARstart{S}{ensor} selection is essential for us to attain scientific and engineering goals with a limited number of sensors, which contributes to saving installation and maintenance costs of the sensors.
Sensors just provide a projection of target information while the information that we want to know can be estimated from the measured information.
Efficient sensor selection enables us to estimate the target information accurately by constructing a reasonable sensor network.
That is useful for detection of chemical gas \cite{legg2012stochastic}, analysis of brain signals \cite{yeo2022efficient}, identification of geolocation \cite{zhang2009fast}, measurement and control of fluid \cite{kaneko2021data,kanda2021feasibility,kanda2022proof,jin2022optimal,inoue2021data,inoue2023datab}, and monitoring of air pollution \cite{kelp2022new}, chip temperature \cite{ranieri2015near}, seismic wavefield reconstruction \cite{nagata2022seismic,nakai2023observation}, and structure health \cite{ostachowicz2019optimization} to list a few.

Sensor-selection problems can be formulated as combinatorial optimization problems, few of which have been proven to be solvable in polynomial time. 
Some sensor-selection problems have been proven NP-hard \cite{natarajan1995sparse, zhang2017sensor}.
The strict optimization often causes fatally long computation time and, therefore, cannot be implemented for a problem with many candidates of sensor positions, especially for a problem in which sensors are placed in two or three dimensional space.
Hence, it is important to develop scalable algorithms that generate good solutions rather than strictly optimal solutions.
Since such an algorithm can be customized depending on the size, the objective function, and the constraint of the problem, a wide variety of algorithms, for example, using cross-entropy optimization \cite{naeem2009cross}, Baysian optimization \cite{garnett2010bayesian}, and branch and bound \cite{chanekar2017optimal}, have been proposed.

Convex relaxation is used to infer solutions to sensor-selection problems and mathematically equivalent actuator-selection problems, e.g. \cite{liu2016sensor, yamada2023efficient}.
Boolean constraints which refer to whether each sensor is selected or not are relaxed to convex box constraints.
If the objective function and the other constraints of the relaxed problem is convex, the strictly optimal solution can be obtained to the relaxed problem with relatively low computational complexity.
Joshi and Boyd proposed an algorithm that solve the relaxed problem by using penalty functions and the Newton method \cite{joshi2009sensor}.
Moreover, the fast algorithms that are modified from the Newton method have been proposed for problems with a large number of sensor candidates \cite{gower2019rsn, nonomura2021randomized}.
Since the optimal solution to the relaxed problem is not feasible for the original problem, except in some special cases as in \cite{romao2018distributed}, the solution is required to be modified so that the boolean constraints are satisfied by rounding. 

Adding a sparse-promoting penalty such as the $0$- and $1$-norms to the objective function is a useful technique for sensor selection.
The sparse-promoting penalization enhances the sparsity of variables after the convex relaxation \cite{summers2016convex, chepuri2015sparsity} and originally continuous variables \cite{lin2013design, munz2014sensor, liu2014energy, dhingra2014admm, nagata2021data, nagata2022data}.
Zare and Jovanovi\'{c} proposed sensor selection for the Kalman filtering by imposing a sparse-promoting penalty on a matrix related to the Kalman gain \cite{zare2018optimala}.
In addition to the Newton method, the proximal gradient and the alternating direction method of multipliers are used to find an optimal solution to this kind of problem.

A greedy method is based on a straightforward concept but often provides a good sensor selection.
Selected sensors are represented by a set of sensor-candidate indices, and the greedy solution is obtained by sequentially adding the optimal index at the time.
The greedy method is applied to sensor/actuator selection customized to linear \cite{jiang2016sensor, silva2019model, manohar2022optimal, takahashi2023sensor} and nonlinear estimation/control \cite{krause2008near, otto2022inadequacy}.
Although the greedy method is heuristic, the performance of the solution can be guaranteed by using properties of the objective function in some problems \cite{shamaiah2010greedy, ranieri2014near, summers2016submodularity, summers2019performance, kohara2020sensor, chamon2021approximate, guo2021actuator}.
For improvement of the performance, the extended greedy heuristics, which store more temporary solutions than the pure greedy method, have been developed in \cite{jiang2019group, nakai2022nondominated, nagata2023randomized,saito2023sensor}.

In addition to a baseline numerical optimization method, whether a sensor selection method is data-driven or not is also an important factor.
A sensor selection method is generally based on a model between quantities to be estimated and measurements.
The underlying model is assumed to be available in advance in a non-data-driven selection method.
Although the underlying model may have been obtained by data-driven modeling, the framework does not include data-driven modeling.
On the other hand, the underlying model is assumed to be not available in advance in a data-driven selection method.
Data from which the underlying model can be identified are available instead, and the framework includes data-driven modeling. 
A moderately complex model is derived to facilitate sensor selection in the framework.
Hence, data-driven sensor selection is beneficial when an underlying model is unknown or a known model is too complex.

Greedy data-driven sensor selection for large-size data has attracted attention recently.
Since a naively-implemented greedy algorithm to sensor selection based on enormous data takes very long time to select sensors, scalable algorithms have been developed.
Manohar~et~al. proposed an algorithm that produces a greedy solution for a sensor selection problem for reconstruction \cite{manohar2018data}.
A linear observation model between latent variables and measured variables is identified by applying the singular value decomposition (SVD) to a data matrix, and pivoting of the QR decomposition for the observation matrix produces the greedy solution.
Their work promoted development of some extended or improved algorithms \cite{clark2019greedy, clark2021multi, saito2021determinant, saito2020data, nakai2021effect, yamada2021fast, saito2021data,yamada2022greedy}.
These algorithms are aimed for sensor selection for estimation of reduced-order quantities to be reconstructed while Li~et~al applied a greedy algorithm \cite{farahat2013efficient} to a sensor selection problem for estimation of the full-order quantities.
The sensor selection algorithm, named as reconstruction-error-oriented efficient greedy (REG) selection, is based on reconstruction by a linear regression in which input quantities are quantities at selected sensor points and target quantities are the quantities at all points.
The efficient algorithm of REG generates a greedy solution to a problem of which objective value is the reconstruction error.
REG outperforms the method of the previous work \cite{manohar2018data} for some datasets in the reconstruction error.

Most data-driven sensor selection methods have focused primarily on reconstruction, in which quantities at all points are estimated from a part of those.
Here, the types of quantities to be estimated and measurement quantities are assumed to be the same.
It is, however, not uncommon to estimate quantities from a different type of the quantities.
For example, the angles and the speed of the relative wind to the aircraft cannot be measured directly, and they are generally estimated from the pressure at sensor locations.
By explicitly setting estimation of target quantities as an objective of a sensor-selection problem, the selected sensors are expected to provide accurate estimation of the target quantities.

We propose generalized REG (GREG), which is an extension of REG \cite{li2021efficient} to a sensor selection method for general estimation.
The target quantities are supposed to be estimated by a ridge-regression estimator which is derived from data.
GREG evaluates the estimation error of the target quantities in the objective function and greedily selects sensors. 
The algorithm is elaborated to compute the greedy solution with low time complexity by using some recurrence relations.

Main contributions of this paper are
\begin{enumerate}
    \item formulation of a sensor selection problem for minimizing a loss function of the ridge regression and
    \item derivation of the efficient greedy algorithm, GREG, for the above problem.
\end{enumerate}
GREG is customized for sensor selection for general estimation unlike REG \cite{li2021efficient} and other data-driven methods \cite{manohar2018data, clark2019greedy, clark2021multi, saito2021determinant, saito2020data, nakai2021effect, yamada2021fast, saito2021data,yamada2022greedy}, which are for reconstruction.
Orthogonal matching pursuit (OMP) can be applied to sensor selection for general estimation although OMP is based on the standard linear regression.
In the standard linear regression, overfitting is likely to occur when the sample (data) size is small.
It has been observed in \cite{inoba2022optimization} that the sensor selection based on the standard linear regression caused low estimation accuracy when overfitting is problematic.
Overfitting generally can be avoided by using the ridge regression.
Thus, we formulate the sensor selection problem for minimizing the loss function of the ridge regression.
The ridge-regression estimator for sensors selected based on this problem is expected to take a low loss value and not to cause the overfitting issue for a moderate regularization parameter.
When the sample size is large, the computation cost for sensor selection is problematic.
To address this, We derive the efficient greedy algorithm for our problem.
GREG requires storing more variables than REG but has the same order of the time computational complexity as REG.
The estimation accuracy and the computation time for GREG and other data-driven sensor-selection algorithms are investigated for some artificial datasets and a real-world dataset.
The effect of the regularization parameter on sensor selection of GREG is also investigated there.

The reminder of this paper is organized as follows:
In Section~\ref{sec: problem}, we formulate the sensor selection problem and explain the computation challenge of the greedy algorithm for that problem.
In Section~\ref{sec: GREG}, we derive the recurrence relations and develop the fast greedy algorithm, GREG.
In Section~\ref{sec: related-work}, we explain related works of data-drive sensor-selection algorithms.
In Section~\ref{sec: application}, we show results of the applications to some datasets with comparison to other sensor-selection algorithms.
Finally, we conclude with a summary of the results in Section~\ref{sec: conclusion}.

\textit{Notation:}
Index sets $\Omega = \{1, 2, \dots, N\}$ and $S = \{S_1, S_2, \dots, S_k\} \subset \Omega$ represent sets of the all sensor candidates and the selected sensors, respectively.
Vector $x_S = [x_{S_1}, x_{S_2}, \dots, x_{S_k}]^{\top} \in \mathbb{R}^k$ has $S_i$th component of vector $x$ in its $i$th component.
$X_S \in \mathbb{R}^{k \times M}$ is a matrix of which $i$th row is equivalent to $S_i$th row of $X$ for each $i \in \{1, 2, \dots, k\}$.
Likewise, $X_{\cdot, S} \in \mathbb{R}^{M \times k}$ is a matrix of which $i$th column is equivalent to $S_i$th column of $X$.
$P_{S, S} \in \mathbb{R}^{k \times k}$ is a matrix of which $(i, j)$ entry is $P_{S_i, S_j}$ for each $i$, $j \in \{1, \dots, k\}$.
We denote a set of consecutive positive integers ranging from $a$ to $b$, $\{a, a+1, \dots, b\} \subset \mathbb{N}$, by $a:b$.
The notation shown in Table~\ref{tb: notation} can be used without description, hereinafter.
\begin{table}[tbh]
  \centering
  \caption{Notation.}
  \label{tb: notation}
  {\renewcommand\arraystretch{1.3}
  \tabcolsep = 3pt
  \begin{tabularx}{\linewidth}{lX}
    \toprule
    Symbol & Description \\
    \midrule
    $\mathbb{R}_{+}$ & $[0, \infty)$, set of nonnegative real numbers \\
    $a:b$ & $\{a, a+1, \dots, b\} \subset \mathbb{N}$, set of consecutive positive integers ranging from $a$ to $b$ \\
    $\Omega$ & $ 1:N $, index set of all sensor candidates \\
    $S$ & $\{S_1, S_2, \dots, S_k\} \subset \Omega$, index set of selected sensors \\
    $x_i$ & $i$th component of vector $x$ \\
    $x_S$ & $[x_{S_1}, x_{S_2}, \dots, x_{S_k}]^{\top}$, vector of which $i$th component is ${S_i}$th component of vector $x$ \\
    $X_{i, j}$ & $(i, j)$ entry of matrix $X$ \\
    $X_{S}$ & matrix of which $i$th row is ${S_i}$th row of matrix $X$ \\
    $X_{\cdot, S}$ & matrix of which $i$th column is ${S_i}$th column of matrix $X$ \\
    $P_{S, S}$ & matrix of which $(i, j)$ entry is $(S_i, S_j)$ entry of square matrix $P$ \\
    $(x(m))_{m = 1}^M $ & $(x(1), x(2), \dots, x(M))$, sequence of $x(m)$ \\
    $\abs{S}$ & cardinal number of set $S$ \\
    $\norm{x}$ & $\sqrt{\sum_{i} x_i^2}$, standard 2-norm of vector $x$ \\
    $\norm{X}_{\mathrm{F}}$ & $\sqrt{\sum_{i} \sum_{j} X_{i, j}^2}$, Frobenius norm of matrix $X$ \\
    $\trace P$ & $\sum_i P_{i, i} $, trace of matrix $P$ \\
    $P \succ 0$ & matrix $P$ is positive definite \\
    \bottomrule
  \end{tabularx}
  }
\end{table}

\section{Sensor Selection for Estimation \label{sec: problem}}
\subsection{Problem setting}
We introduce a sensor selection problem based on subset selection in the ridge regression.
The objective is to select sensors so that the target variables can be estimated from the selected measurements with high accuracy.
We assume that snapshot pairs of the target quantities to be estimated and the quantities that can be measured by all sensors are available for sensor selection. 
Although these snapshot data would be obtained by simulations or experiments, any model between the target and measurement quantities is not required beforehand.
Here, the model is inferred from the snapshot data by the ridge regression, and the sensors are selected so that the loss value of the regression would be minimized. 

Let us denote the snapshots of the target vector and the all-measurement vector at instance $m \in \mathbb{N}$ by $y(m) \in \mathbb{R}^{N_y}$ and $x(m) \in \mathbb{R}^{N}$, respectively, and denote the number of the snapshot pairs by $M \in \mathbb{N}$. 
Each element $x_i(m)$ of $x(m)$ represents the measurement value of the $i$th sensor.
If we denote the selected sensors by $S = \{S_1, S_2, \dots, S_k\} \subset \Omega$, quantities measured with the selected sensors $S$ are expressed by $x_S(m) \in \mathbb{R}^k$.
Our objective is to select sensors $S$ so that a new target vector $y^{\mathrm{new}} \in \mathbb{R}^{N_y}$ can be accurately estimated from a new input vector $x_S^{\mathrm{new}} \in \mathbb{R}^k$ that is measured with sensors $S$.

In the ridge regression, the new unknown target $y^{\mathrm{new}}$ is estimated as follows:
\begin{align}
  y^{\mathrm{new}} = K \, x_S^{\mathrm{new}} + \epsilon,
  \label{eq: linear_regression_model}
\end{align}
where $K \in \mathbb{R}^{N_y \times k}$ is the matrix determined based on the training data $(x(m), y(m))_{m = 1}^M$, and $\epsilon \in \mathbb{R}^{N_y}$ is the estimation error vector. 
In the estimation system, the estimate of $y^{\mathrm{new}}$ can be obtained by multiplying the new measurements $x_S^{\mathrm{new}}$ with the selected sensors and the gain matrix $K$.
The objective is to construct an estimation system that predicts the new unknown targets with a small estimation error.
Here, the estimation matrix $K$ is determined based on minimization of the following loss function:
\begin{align}
  J_0(K, S) = \frac{1}{M} \sum_{m = 1}^M \norm{y{(m)} - K \, x_S{(m)}}^2
  + \tilde{\lambda} \norm{K}_{\mathrm{F}}^2, \label{eq: cost0}
\end{align}
where $\tilde{\lambda} \in \mathbb{R}_+$ is a regularization parameter.
The first term of the loss function $J_0$ is the estimation error for the training data, and the second term is the regularization one.
Minimizing only the first term of the training error does not always yield good estimation models.
Overfitted models that estimate the targets in the training data extremely well often fail to estimate the unknown targets \cite{bishop2006pattern}. 
The overfitted linear models tend to have large magnitude of $K$, and the regularization term serves to prevent overfitting by suppressing the magnitude of $K$.
The models with small loss value for a appropriate regularization parameter are expected to estimate the unknown targets accurately.

Since the input variables, measurements of the selected sensors, depend on the selected sensors set $S$, the loss function \eqref{eq: cost0} depends not only on $K$ and but also on $S$ unlike that of the conventional ridge regression.
Thus, there is room for reducing the loss by optimizing the sensor set $S$.

For fixed $S$, the gain minimizing $J_0$ is given by
\begin{align}
  K^*(S) = Y \, X_S^{\top} (X_S X_S^{\top} + \lambda I_{\abs{S}})^{-1},
  \label{eq: Kopt}
\end{align}
where $\lambda = M \, \tilde{\lambda}$, $X = [x(1), x(2), \dots, x(M)]$, $Y = [y(1), y(2), \dots, y(M)]$, and $X_S X_S^{\top} + \lambda I_{\abs{S}}$ is assumed to be positive definite.
The substitution $K = K^*(S)$ in (\ref{eq: cost0}) leads to
\begin{align}
  J_0(K^*(S), S) = & \frac{1}{M} \trace \Bigl\{
    Y Y^\top \notag \\
    & - Y \, X_S^{\top} (X_S X_S^{\top} + \lambda I_{\abs{S}})^{-1} X_S Y^\top
  \Bigr\}.
  \label{eq: cost0_opt}
\end{align}
Equation~\eqref{eq: cost0_opt} expresses the minimized loss function with respect to $K$ for fixed $S$.
The partially minimized loss function $J_0(K^*(S), S)$ can be further reduced by optimizing the sensor set $S$ to minimize the original loss function $J_0(K, S)$.
Hence, we consider the following sensor selection problem:
\begin{problem} \label{pb: selection_LR}
  Given $X \in \mathbb{R}^{N \times M}$, $Y \in \mathbb{R}^{N_y \times M}$, $\lambda \in \mathbb{R}_+$, and $p \in \mathbb{N}$ with $p \le N$,
  find $S \subset \Omega$ that maximizes
  \begin{align}
    J(S) = \trace \left\{
     Y X_S^{\top} (X_S X_S^{\top} + \lambda I_{\abs{S}})^{-1} X_S Y^\top
      \right\}
    \label{eq: cost}
  \end{align}
  subject to
  \begin{align}
    & X_S X_S^{\top} + \lambda I_{\abs{S}} \succ 0,
    \label{eq: constraint1} \\
    & \abs{S} \le p.
    \label{eq: constraint2}
  \end{align}
\end{problem}
Note that a solution maximizing $J(S)$ minimizes the cost $J_0(K^*(S), S)$ because $Y Y^\top$ in \eqref{eq: cost0_opt} is constant.
The constraint (\ref{eq: constraint1}) in Problem~\ref{pb: selection_LR} is imposed for the existence of the inverse in the objective function $J$.
The constraint (\ref{eq: constraint1}) is inactive for $\lambda > 0$.
On the other hand, (\ref{eq: constraint1}) is active for $\lambda = 0$, and $X_S$ must be of full rank to satisfy \eqref{eq: constraint1}.
The constraint \eqref{eq: constraint2} limits the number of the sensors, which is assumed to be much smaller than that of the sensor candidates, i.e, $p \ll N$.
Problem~\ref{pb: selection_LR} is aimed to select sensors so that the loss \eqref{eq: cost0_opt} in the ridge regression is as small as possible.
Hence, sensor selection based on Problem~\ref{pb: selection_LR} yields a sensor set which constructs the estimation system \eqref{eq: linear_regression_model} without overfitting and with high estimation accuracy.
GREG, which is developed in this paper, efficiently computes a suboptimal sensor set to Problem~\ref{pb: selection_LR}.


REG \cite{li2021efficient}, which is sensor selection for reconstruction, is based on a linear model in which the target vector $y$ in \eqref{eq: linear_regression_model} is replaced by the all-measurement vector $x$. 
The linear model in REG is derived by the standard linear regression, which is the ridge regression under a special condition of $\lambda = 0$.

The arbitrary nature of the quantities estimated in the model separates sensor selection for general estimation from sensor selection for reconstruction.
In sensor selection for reconstruction, as in REG, the quantities to be estimated are restricted to the all-measurement quantities or their reduced-order ones. 
However, there is no such restriction in sensor selection for general estimation in GREG, in which any continuous quantity can be set as the target quantity.
Section~\ref{sec: related-work} presents details of estimation models in other data-driven sensor-selection algorithms.

Nonlinear regression with kernel functions and neural networks has been widely used for data-driven modeling between measured variables and target variables, e.g. in \cite{yoon2019gaussian, tian2022flexible}.
Kernel functions and neural networks have universal approximation properties under certain conditions \cite{micchelli2006universal, hornik1991approximation}, and regression using these nonlinear functions generally provides a more accurate model than ridge regression.
However, the simple formation of the ridge regression makes the sensor selection problem not complicated.
This facilitates the derivation of an efficient optimization algorithm.

\subsection{Greedy selection}

The optimal solution to Problem~\ref{pb: selection_LR} cannot be found by brute-force search or branch and bound within realistic time unless the number $N$ of sensor candidates is quite few due to combinatorial explosion.
Hence, we find a suboptimal solution with the use of the greedy algorithm.
The greedy algorithm adds an element that causes a maximum increase in an objective function to a temporary set at each iteration and finally returns the set with a desirable cardinal number.

\begin{algorithm}[t]
  \caption{Greedy Algorithm} \label{alg: greedy}
  \begin{algorithmic}[1]
      \STATE \textbf{Input:} $X \in \mathbb{R}^{N \times M}$, $Y \in \mathbb{R}^{N_y \times M}$,
      $p \in \mathbb{N}$, $\tilde{\lambda} \in \mathbb{R}_{+}$
      \STATE Initialize $\lambda = M \tilde{\lambda}$, \, $S = \emptyset$
      \FOR{$k = 1, 2, \dots, p$}
      \STATE $\mathcal{I} = \set{i \in \Omega \setminus S}{S' = S \cup \{i\}, \, X_{S'} X_{S'}^{\top} + \lambda I_{\abs{S'}} \succ 0}$
      \STATE $s \in \argmax_{i \in \mathcal{I}} \left\{ J(S \cup \{i\}) - J(S) \right\}$
      \STATE $S \leftarrow S \cup \{ s \}$
      \ENDFOR
      \RETURN $S$
  \end{algorithmic}
\end{algorithm}

Algorithm~\ref{alg: greedy} shows the greedy algorithm for Problem~\ref{pb: selection_LR}.
In Algorithm~\ref{alg: greedy}, the temporary set is initialized as $S = \emptyset$ at first.
Then, a set of elements $i$ that construct feasible solutions $S \cup \{i\}$ satisfying the constraint \eqref{eq: constraint1} is identified as $\mathcal{I}$.
An element that maximizes an increase $J(S \cup \{i\}) - J(S)$ in an objective value among all feasible elements $\mathcal{I}$ is added to the temporary set as $S \leftarrow S \cup \{ s \}$.
This addition of the element is repeated $p$ times, and consequently the greedy solution $S$ with $\abs{S} = p$ is obtained.

All of the greedy solution and their temporary solutions are expressed as $S$ in the pseudocode due to the simplicity of the notations. 
We distinguish these notations by the iteration number $k$ for more detailed discussions out of the pseudocode hereinafter.
The greedy element $\bar{s}_k \in \Omega$ and solution $\bar{S}_k \subset \Omega$ at the $k$th iteration in the greedy algorithm are recursively defined as
\begin{align}
  & \bar{s}_k \in \argmax_{i \in \mathcal{I}_k} \left\{ J(\bar{S}_{k-1} \cup \{i\}) - J(\bar{S}_{k-1}) \right\}, \notag \\
  & \bar{S}_k = \bar{S}_{k-1} \cup \left\{\bar{s}_k\right\} \notag 
\end{align}
for $k = 1, 2, \dots, p$ with start from $\bar{S}_0 = \emptyset$, where
\begin{align}
  \mathcal{I}_k = \bigl\{ i \in \Omega \setminus \bar{S}_{k-1} \mid
  & S' = \bar{S}_{k-1} \cup \{i\}, \notag \\
  & X_{S'} X_{S'}^{\top} + \lambda I_{\abs{S'}} \succ 0 \bigr\}
  \label{eq: feasible_set}
\end{align}
is a set of feasible elements with which $S' = \bar{S}_{k-1} \cup \{i\}$ satisfies the constraint \eqref{eq: constraint1}.

Algorithm~\ref{alg: greedy} with $\lambda = 0$ appears in the field of machine learning as a feature selection algorithm for standard linear regression. 
The feature selection according to algorithm~\ref{alg: greedy} is called as forward regression or forward selection \cite{miller2002subset}.

Some works have tackled with performance guarantees of the greedy solutions of the forward selection.
Das~et~al. showed that a performance guarantee of the forward selection that is similar to that of OMP \cite{pati1993orthogonal, tropp2004greed} in a case in which nondiagonal entries of a covariance matrix of independent variables are close to zero \cite{das2008algorithms}.
They also showed the objective function in the subset selection for the linear regression is not submodular in general.
The lack of the submodularity means that the well-known Nemhausers' results \cite{nemhauser1978analysis, krause2014submodular} cannot be employed for the performance guarantee.
Elenverg~et~al. derived a guarantee that does not need any strong condition with respect to dependent and independent variables by focusing on restricted strong concavity and restricted smoothness of the objective function \cite{elenberg2018restricted}.
Das~et~al. also gave a general guarantee that can be practically computed by using a submodularity ratio that is a relaxed submodualrity and derived a lower bound for the greedy solution in a case of a single target variable in \cite{das2011submodular, das2018approximate}.
The same lower bound can be derived even in a case of multiple target variables in a similar manner to \cite{das2011submodular, das2018approximate}, and it gives the performance guarantee of Algorithm~\ref{alg: greedy}.
%

This paper focuses on the implementation of the greedy algorithm for Problem~\ref{pb: selection_LR}, i.e., Algorithm~\ref{alg: greedy}.
Quite a long time is required to compute a greedy solution for large data by the naive implementation, where $J(\bar{S}_{k-1} \cup \{i\})$ for all $i \in \mathcal{I}_k$ are computed to choose the optimal element in each iteration.
REG, in which a forward-selection algorithm in \cite{farahat2013efficient} is repurposed, is an efficient implementation of Algorithm~\ref{alg: greedy} in the reconstruction case $(Y, \lambda) = (X, 0)$ but is not applicable in the general case $(Y, \lambda) \neq (X, 0)$.
We will develop another efficient implementation of Algorithm~\ref{alg: greedy}, GREG, for the general case in the next section.

\section{Fast Algorithm for Greedy Selection \label{sec: GREG}}
The long computation time of the naive implementation is due to the matrix products and inversion in the objective function \eqref{eq: cost}, which cause the third-order time complexity with respect to $N$, $N_y$, $M$, and $p$.
Since the objective function is evaluated $N \times p$ times for selecting $p$ sensors, the naive implementation has the fifth-order complexity as shown in Table~\ref{tb: complexity}.
The high complexity of the naive implementation is because matrix multiplications and matrix inversion must be computed for evaluating $J(S \cup \{i\})$ for each $i$ based on \eqref{eq: cost}.
We will develop the efficient implementation with the third-order complexity by deriving two sets of the difference equations with respect to the iteration number $k$.
The first set is a difference equation of the objective function $J$ for the greedy solutions.
The second set is a pair of difference equations of two variables in the first difference equation.
In the developed implementation, matrix multiplications and matrix inversion are not required at each iteration.


\begin{table}[!t]
  \centering
  \caption{Computational complexity.}
  \label{tb: complexity}
  \begin{tabular}{ll}
    \toprule
    Algorithm & Computational complexity \\
    \midrule
    Naive & $O(N p^4 + N M p^3 + N N_y M p^2 + N N_y p^3)$ \\
    GREG & $O(N_y M (\min \{ N_y, M \} + N) $ \\
         & $\quad\quad\quad + (N + N_y) M p + (N + N_y) p^2)$ \\
    REG \cite{li2021efficient} & $O(N M \min \{ N, M \} + N M p + N p^2)$ \\
    SOMP \cite{tropp2006algorithms} & $O(N N_y M + (N N_y + N M + N_y M) p + M p^2)$ \\
    DG \cite{saito2021determinant} & 
    $\begin{array}{ll}
      O(N r^2) & p \le r \\
      O(N r^2 p) & p > r
    \end{array}$\\
    BDG \cite{yamada2021fast} & $O(N r^2 p + (N M + N r) p^2 + r p^3)$ \\
    \bottomrule
  \end{tabular}
\end{table}

The difference equations for the efficient implementation will be derived by utilizing the fact that the objective function can be written with some covariance-like matrices.
Let us define the covariance-like matrices
$P^{x x} \in \mathbb{R}^{N \times N}$,
$P^{x y} \in \mathbb{R}^{N \times N_y}$,
$P^{y y} \in \mathbb{R}^{N_y \times N_y}$,
$Q^{x x}(S) \in \mathbb{R}^{N \times N}$,
$Q^{x y}(S) \in \mathbb{R}^{N \times N_y}$, and
$Q^{y y}(S) \in \mathbb{R}^{N_y \times N_y}$ as follows:
\begin{align}
  &
  P^{x x} = X X^\top + \lambda I_{N}, \,
  P^{x y} = X Y^\top, \,
  P^{y y} = Y Y^\top, \\
  &
  Q^{x x}(\emptyset) = P^{x x}, \,
  Q^{x y}(\emptyset) = P^{x y}, \,
  Q^{y y}(\emptyset) = P^{y y},
  \label{eq: init_Q} \\
  & 
  Q^{x x}(S) = P^{x x} - (P^{x x}_S)^\top (P^{x x}_{S, S})^{-1} P^{x x}_S, \,
  \forall S: \abs{S} \ge 1,
  \label{eq: Qxx} \\
  & 
  Q^{x y}(S) = P^{x y} - (P^{x x}_S)^\top (P^{x x}_{S, S})^{-1} P^{x y}_S, \,
  \forall S: \abs{S} \ge 1,
  \label{eq: Qxy} \\ & 
  Q^{y y}(S) = P^{y y} - (P^{x y}_S)^\top (P^{x x}_{S, S})^{-1} P^{x y}_S, \,
  \forall S: \abs{S} \ge 1,
  \label{eq: Qyy}
\end{align}
for $S \subset \Omega$ that satisfies the positive-definiteness constraint \eqref{eq: constraint1}, i.e., $P^{x x}_{S, S} = X_S X_S^{\top} + \lambda I_{\abs{S}} \succ 0$.
Matrices $P^{x x}$, $P^{x y}$, and $P^{y y}$ are the sampling prior covariance matrices with respect to $(x, x)$,  $(x, y)$, and  $(y, y)$ with some scaling parameters, respectively, and $Q^{x x}(S)$, $Q^{x y}(S)$, and $Q^{y y}(S)$ are their posterior ones given $x_S$.
The difference in $J$ with adding $i$ is expressed by the difference in $Q_{yy}$ as shown in Appendix~\ref{sec: pf_incJ}.
By using this fact, a difference equation of $J$, which can be evaluated with low computational complexity, is derived:
\begin{theorem}
  \label{thm: incJ}
  For any $S \subset \Omega$ and $i \in \Omega \setminus S$ so that $P^{xx}_{S \cup \{i\}, S \cup \{i\}} \succ 0$, the following difference equation holds:
  \begin{align}
    & J(S \cup \{ i \}) - J(S)
  = \frac{\Norm{(Q^{x y}_i (S))^{\top}}^2}{Q^{x x}_{i, i} (S)}.
  \label{eq: update_J}
  \end{align}
\end{theorem}
\begin{proof}
    See Appendix~\ref{sec: pf_incJ} for the proof.
\end{proof}
The difference equation \eqref{eq: update_J} does not include either matrix product or matrix inversion explicitly.
Hence, the increase $J(S \cup \{ i \}) - J(S)$ can be computed with low computational complexity if the vector $Q^{x y}_i (S)$ and the scalar $Q^{x x}_{i, i} (S)$ are given.

The difference equation \eqref{eq: update_J} with $S = \bar{S}_{k-1}$ can be used at the $k$th iteration of the greedy algorithm.
The increase $J(S \cup \{ i \}) - J(S)$ is computed for each $i \in \mathcal{I}_{k}$ with \eqref{eq: update_J}, and the optimal $i$ is selected.
The above implementation of the greedy algorithm is a prototype version of GREG.
Although the prototype implementation is faster than the naive one, the computational complexity is still forth order even if the computational cost for $\mathcal{I}_k$ is ignored.
A little high complexity of the prototype implementation comes from the cost for computing $Q^{x y}_{i} (\bar{S}_{k-1})$ and  $Q^{x x}_{i, i} (\bar{S}_{k-1})$ for all $i \in \mathcal{I}_k$.
According to \eqref{eq: Qxy} and \eqref{eq: Qxx}, computations with matrix inversion $(P^{x x}_{\bar{S}_{k-1}, \bar{S}_{k-1}})^{-1}$ with the third-order complexity are required to get $Q^{x y}_{i} (\bar{S}_{k-1})$ and  $Q^{x x}_{i, i} (\bar{S}_{k-1})$ for each $k \in 1:p$.

For further reduction in the computational complexity of the greedy algorithm, we derive efficient difference equations of the numerator $\Norm{Q^{x y}_i}(S)$ and the denominator $Q^{x x}_{i, i}(S)$ in \eqref{eq: update_J}.
The difference of $J$ in \eqref{eq: update_J} is evaluated via the difference equations of $\Norm{Q^{x y}_i}(S)$ and $Q^{x x}_{i, i}(S)$ instead of directly computing $Q^{x y}_i(S)$ and $Q^{x x}_{i, i}(S)$.
Let us denote the numerator and the denominator in (\ref{eq: update_J}) by $f_i^{(k)}$ and $g_i^{(k)}$, respectively, as follows:
\begin{align}
  f_i^{(k)} & \coloneqq \Norm{(Q^{x y}_i (\bar{S}_k))^{\top}}^2,
  \label{eq: f} \\
  g_i^{(k)} & \coloneqq Q^{x x}_{i, i} (\bar{S}_k).
  \label{eq: g} 
\end{align}
At first of GREG, $f_i^{(k)}$ and $g_i^{(k)}$ for all feasible elements $i \in \mathcal{I}_k$ are initialized via the definition equations of $Q^{x y}(\emptyset)$ and $Q^{x x}(\emptyset)$ with $k = 1$.
Next, the objective-value increase $J(\bar{S}_k \cup \{ i \}) - J(\bar{S}_k) = f_i^{(k)} / g_i^{(k)}$ is computed for any $i \in \mathcal{I}_k$.
The optimal element $s$, which yields the maximum increase, is selected as the first greedy element.
Then, the optimal element is added to the temporal solution as $\bar{S}_{k+1} = \bar{S}_k \cup \{s\}$, and $f_i^{(k+1)}$ and $g_i^{(k+1)}$ are updated via the difference equations which we derive later.
The procedure of computing the objective-value increase, selecting the optimal element, and updating $f_i$ and $g_i$ is done $p - 1$ times in addition, and finally the greedy solution with the cardinal number $p$ can be obtained.
This implementation of the greedy algorithm has only the third-order time complexity due to avoiding direct computation of $Q^{x y}$ and $Q^{x x}$.


We derive difference equations related to the decrements in $Q^{xx}$ and $Q^{xy}$ for deriving update formulas of $f_i$ and $g_i$.
We introduce vectors $\xi^{(k)}$ and $\theta^{(k)}$ that constructs rank-one matrices as follows:
\begin{align}
  \xi^{(k)} {\xi^{(k)}}^{\top} & = (Q^{x x}_{\bar{s}_k} (\bar{S}_{k-1}))^{\top} (Q^{x x}_{\bar{s}_k, \bar{s}_k} (\bar{S}_{k-1}))^{-1} Q^{x x}_{\bar{s}_k} (\bar{S}_{k-1}),
  \notag \\
  \xi^{(k)} {\theta^{(k)}}^{\top} & = (Q^{x x}_{\bar{s}_k} (\bar{S}_{k-1}))^{\top} (Q^{x x}_{\bar{s}_k, \bar{s}_k} (\bar{S}_{k-1}))^{-1} Q^{x y}_{\bar{s}_k} (\bar{S}_{k-1}).
  \notag
\end{align}
Formal definitions of $\xi^{(k)}$ and $\theta^{(k)}$ will be given in later Lemma~\ref{lm: xi_theta}.
Vectors $\xi^{(k)}$ and $\theta^{(k)}$ become keys of the update formulas of $f_i$ and $g_i$.
The rank-one matrices constructed by $\xi^{(k)}$ and $\theta^{(k)}$ are the decrements in $Q^{x x}$ and $Q^{x y}$ at the $k$th iteration in the greedy algorithm.
Hence, $Q^{x x}$ and $Q^{x y}$ can be expressed by using present and past $\xi$ and $\theta$.
The difference equations of $f$ and $g$ using $\xi$ and $\theta$ will be derived by means of this property.
The following lemma gives the update formulas of $\xi$ and $\theta$ and the expressions of $Q^{x x}$ and $Q^{x y}$.
\begin{lemma}
  \label{lm: xi_theta}
  Define $\delta^{(k)} \in \mathbb{R}^{N}$, $\xi^{(k)} \in \mathbb{R}^{N}$, $\theta^{(k)} \in \mathbb{R}^{N_y}$, $\Xi^{(k)} \in \mathbb{R}^{N \times k}$, and $\Theta^{(k)} \in \mathbb{R}^{N_y \times k}$ recursively from $k = 1$ to $k = p$ with initial settings $\Xi^{(0)} \in \mathbb{R}^{N \times 0}$ and $\Theta^{(0)} \in \mathbb{R}^{N_y \times 0}$ by the following equations:
  \begin{align}
    &
    \delta^{(k)} = X X_{\bar{s}_k}^{\top} + \lambda \, (I_{N})_{\bar{s}_k}^{\top} - \Xi^{(k-1)} (\Xi^{(k-1)}_{\bar{s}_k})^{\top}, \label{eq: delta} \\
    &
    \xi^{(k)} = \delta^{(k)} / \sqrt{\delta^{(k)}_{\bar{s}_k}}, \label{eq: omega} \\
    &
    \theta^{(k)} = (Y X_{\bar{s}_k}^{\top} - \Theta^{(k-1)} (\Xi^{(k-1)}_{\bar{s}_k})^{\top}) / \sqrt{\delta^{(k)}_{\bar{s}_k}}, \label{eq: theta} \\
    &
    \Xi^{(k)} = [\Xi^{(k-1)}, \, \xi^{(k)}], \quad
    \Theta^{(k)} = [\Theta^{(k-1)}, \, \theta^{(k)}],
    \label{eq: Xi_Theta}
  \end{align}
  where $(I_{N})_{\bar{s}_k}^{\top}$ is a canonical unit vector of which $\bar{s}_k$th component is $1$, and, for empty matrices $\Xi^{(0)}$ and $\Theta^{(0)}$, 
  \begin{align}
    &
    \Xi^{(0)} (\Xi^{(0)}_{\bar{s}_1})^{\top} = 0, \, \,
    \Theta^{(0)} (\Xi^{(0)}_{\bar{s}_1})^{\top} = 0,
    \notag \\
    &
    [\Xi^{(0)}, \, \xi^{(1)}] = \xi^{(1)}, \, \,
    [\Theta^{(0)}, \, \theta^{(1)}] = \theta^{(1)}.
    \notag
  \end{align}
  For any $k \in 1:p$, the following equations hold:
  \begin{align}
    \xi^{(k)} & = {Q_{\bar{s}_k}^{xx}(\bar{S}_{k-1})}^{\top} ({Q_{\bar{s}_k, \bar{s}_k}^{xx}(\bar{S}_{k-1})})^{-1/2},
    \label{eq: omega2} \\
    \theta^{(k)} & = {Q_{\bar{s}_k}^{xy}(\bar{S}_{k-1})}^{\top} ({Q_{\bar{s}_k, \bar{s}_k}^{xx}(\bar{S}_{k-1})})^{-1/2},
    \label{eq: theta2} \\
    Q^{x x}(\bar{S}_{k}) & = Q^{x x}(\bar{S}_{k-1}) - \xi^{(k)} {\xi^{(k)}}^{\top}
    \label{eq: incQxx2} \\
    & = X X^{\top} + \lambda I_{N} - \Xi^{(k)} {\Xi^{(k)}}^{\top},
    \label{eq: incQxx3} \\
    Q^{x y}(\bar{S}_{k}) & = Q^{x y}(\bar{S}_{k-1}) - \xi^{(k)} {\theta^{(k)}}^{\top}
    \label{eq: incQxy2} \\
    & = X Y^{\top} - \Xi^{(k)} {\Theta^{(k)}}^{\top}.
    \label{eq: incQxy3}
  \end{align}
\end{lemma}
\begin{proof}
See Appendix~\ref{sec: xi_theta} for the proof.
\end{proof}
Matrices $\Xi^{(k)}$ and $\Theta^{(k)}$ in \eqref{eq: Xi_Theta} are the data matrices of the present and past $\xi$ and $\theta$ at the $k$th iteration.
Vectors $\xi^{(k)}$ and $\theta^{(k)}$ can be given by using the past data matrices $\Xi^{(k-1)}$ and $\Theta^{(k-1)}$ according to \eqref{eq: delta}--\eqref{eq: theta}.
Hence, $\xi^{(k)}$ and $\theta^{(k)}$ can be computed recursively from $k = 1$ to $k = p$.
Equations~\eqref{eq: incQxx2} and \eqref{eq: incQxy2} show that $\xi^{(k)}$ and $\theta^{(k)}$ form the decrements in $Q^{x x}$ and $Q^{x y}$ at the $k$th iteration in the greedy algorithm.
Matrices $Q^{x x}(\bar{S}_{k})$ and $Q^{x y}(\bar{S}_{k})$ can be expressed by using $\Xi^{(k)}$ and $\Theta^{(k)}$ as in \eqref{eq: incQxx3} and \eqref{eq: incQxy3}.

The difference equations of $f_i^{(k)}$ and $g_i^{(k)}$ are derived from the expressions \eqref{eq: incQxx2}, \eqref{eq: incQxy2}, and \eqref{eq: incQxy3} of $Q^{x y}(\bar{S}_{k})$ and $Q^{x x}(\bar{S}_{k})$.
\begin{theorem}
  \label{thm: update_fg}
  For any $k \in 1:p$ and $i \in \mathcal{I}_k$, the following difference equations hold:
  \begin{align}
    f_i^{(k)} & = f_i^{(k-1)} - \xi_i^{(k)} \bigl( 2 X_i Y^{\top} \theta^{(k)} - 2 \Xi_i^{(k-1)} {\Theta^{(k-1)}}^{\top} \theta^{(k)} \notag \\
    & \hspace{4cm} - \Norm{\theta^{(k)}}^2 \xi_i^{(k)} \bigr),
    \label{eq: incf} \\
    g_i^{(k)} & = g_i^{(k-1)} - \xi_i^{(k)} \xi_i^{(k)}.
    \label{eq: incg}
  \end{align} 
\end{theorem}
\begin{proof}
    See Appendix~\ref{sec: update_fg} for the proof.
\end{proof}
The greedy algorithm can be developed by using the difference equations \eqref{eq: incf} and \eqref{eq: incg} of $f_i$ and $g_i$.
The proposed implementation, GREG, is summarized in Algorithm~\ref{alg: EEG}.
Here, the overbar for the greedy solution and the index $k$ are omitted for simplicity of the notations.
The inputs to GREG are the data matrices $X$ and $Y$, the number $p$ of the sensors, and the regularization parameter $\tilde{\lambda}$.
In the first process, $f_i$ and $g_i$ are initialized based on the definition equations \eqref{eq: f} and \eqref{eq: g} for each $i \in \Omega$  on line~4, and the iteration starts with $k = 1$.
Then, the feasible set $\mathcal{I}$ is identified based on the values of $g$ on line~6.
See Appendix~\ref{sec: identification} for details of the relationship of $\mathcal{I}$ and $g$.
Next, the optimal element $s$ that maximizes the increment in the objective function, $f_i / g_i$, is selected and added to the temporary greedy solution $S$ on line~7.
Vectors $\xi$ and $\theta$ are computed via \eqref{eq: delta}--\eqref{eq: theta} on lines~9 and 10, and $f$ and $g$ are updated via \eqref{eq: incf} and \eqref{eq: incg} for the next iteration on lines~11 and 12.
Then, $\xi$ and $\theta$ are stored on line~13, and the next iteration starts as $k \leftarrow k + 1$.
The greedy solution is returned after $p$-elements selection.

\begin{algorithm}[b]
  \caption{GREG} \label{alg: EEG}
  \begin{algorithmic}[1]
      \STATE \textbf{Input:} $X \in \mathbb{R}^{N \times M}$, $Y \in \mathbb{R}^{N_y \times M}$,
      $p \in \mathbb{N}$, $\tilde{\lambda} \in [0, \infty)$
      \STATE Initialize $\lambda = M \tilde{\lambda}$, \, $\Xi \in \mathbb{R}^{N \times 0}$, \, $\Theta \in \mathbb{R}^{N_y \times 0}$
      \STATE Initialize $S = \emptyset$, \, $\mathcal{I} = \Omega$
      \STATE Initialize $f_i = \norm{Y X_i^{\top}}^2$, \,
      $g_i = X_{i} X_{i}^{\top} + \lambda$ \, $\forall i \in \mathcal{I}$
      \FOR{$k = 1, 2, \dots, p$}
      \STATE $\mathcal{I} \leftarrow \set{i \in \mathcal{I} \setminus S}{g_i > 0}$
      \STATE $s \in \argmax_{i \in \mathcal{I}} f_i / g_i$,
      \, $S \leftarrow S \cup \{ s \}$
      \STATE \textbf{if} $\abs{S} = p$ \textbf{then};  \textbf{return} $S$;  \textbf{endif}
      \STATE $\delta = X X_{s}^{\top} + \lambda (I_N)_s - \Xi \, \Xi_{s}^{\top}$
      \STATE $\xi = \delta / \sqrt{\delta_s}$, \, \,
      $\theta = (Y X_{s}^{\top} - \Theta (\Xi_{s})^{\top}) / \sqrt{\delta_{s}}$
      \STATE $f_i \leftarrow f_i - \xi_i \bigl( 2 X_i Y^{\top} \theta - 2 \Xi_i {\Theta}^{\top} \theta - \Norm{\theta}^2 \xi_i \bigr)$,
      $\forall i \in \mathcal{I}$
      \STATE $g_i \leftarrow g_i - \xi_i^2$, \,
      $\forall i \in \mathcal{I}$
      \STATE $\Xi \leftarrow [\Xi, \, \xi]$, \, $\Theta \leftarrow [\Theta, \, \theta]$
      \ENDFOR
  \end{algorithmic}
\end{algorithm}

The computational complexity is reduced by using the difference equations of the numerator and the denominator in \eqref{eq: update_J}.
The prototype greedy algorithm, in which $Q_{x x}$ and $Q_{x y}$ are evaluated with \eqref{eq: Qxx} and \eqref{eq: Qxy} for the numerator and the denominator, needs the matrix product and matrix inversion operations in the iteration.
On the other hand, GREG does not need these operations.
Equations~\eqref{eq: delta}--\eqref{eq: theta}, \eqref{eq: incf}, and \eqref{eq: incg} are evaluated in each iteration and only needs the operations of the second-order complexity such as the matrix-vector products.
Hence, the total computational complexity through the $p$ iterations is of third order as shown in Table~\ref{tb: complexity}.


The computational complexity is reduced by the second order from that of the naive algorithm instead of paying the additional memory cost for the $N \times (p-1)$ matrix $\Xi^{(p-1)}$ and the $N_y \times (p-1)$ matrix $\Theta^{(p-1)}$.
This additional memory is much less than that of the $N \times M$ matrix $X$ and the $N_y \times M$ matrix $Y$ when the sampling number $M$ of the data matrices is much less than the number $p$ of the sensors.

\section{Related Works \label{sec: related-work}}
Numerous algorithms have been proposed for data-driven sensor selection or subset selection for accurate linear estimation.
This section provides comparison of GREG with some of the other algorithms.
The sensor-selection algorithms introduced in this section will be applied to two types of datasets in the next section, as well as GREG.
The characteristics of the algorithms are summarized in Table~\ref{tb: characteristics}.
In Table~\ref{tb: characteristics} we call that a greedy method is a direct type if it selects an element maximizing (or minimizing) the objective function at each iteration.
REG, DG, and BDG are designed for estimating the all-measurement vector or its reduced-order one while GREG and SOMP are designed for estimating a vector which can be set arbitrarily.
Hence, REG, DG, and BDG are categorized into sensor-selection algorithms for reconstruction while GREG and SOMP are categorized into sensor-selection algorithms for general estimation.

\begin{table*}[t]
 \centering
 \caption{Characteristics of data-driven sensor-selection algorithms}
 \label{tb: characteristics}
 \begin{tabular}{lccccc}
   \toprule
   Algorithm & Variables to be estimated & Underlying model & Regularization & Model of truncated modes & Type of greedy method \\
   \midrule
   GREG & Arbitrary target vector $y$ & Linear regression model & 2-norm & $-$ & Direct type \\
   REG \cite{li2021efficient} & All-measurement vector $x$ & Linear regression model & Not considered & $-$ & Direct type \\
   SOMP \cite{tropp2006algorithms} & Arbitrary target vector $y$ & Linear regression model & Not considered & $-$ & Indirect type \\
   DG \cite{saito2021determinant} & Reduced-order vector of $x$ & SVD model & $-$ & Uncorrelated noise model & Direct type \\
   BDG \cite{yamada2021fast} & Reduced-order vector of $x$ & SVD model & $-$ & Correlated noise model & Direct type \\
   \bottomrule
 \end{tabular}
\end{table*}

\subsection{REG}
REG specializes in the case of $(Y, \lambda) = (X, 0)$ in Problem~\ref{pb: selection_LR} \cite{li2021efficient}.
The linear regression model \eqref{eq: linear_regression_model} with $y = x$ is used in REG.
Hence, REG is not designed for sensor selection for general estimation in which $y \neq x$ and therefore is not expected to work well for this type of sensor selection.
The difference of the underlying problems in REG and GREG results in the difference of the algorithms.
GREG in Algorithm~\ref{alg: EEG} requires computing and storing $\theta$ for updating $\Norm{Q_i^{x y}}^2$ while REG does not need $\theta$.
Therefore, REG is more efficient than GREG in sensor selection for reconstruction, in which $y = x$.
In addition, the regularization parameter appears in the second term in \eqref{eq: delta} in GREG while the regularization parameter does not appear in REG.
The overfitting of the estimator tends to be problematic when the sample size is small.
GREG is expected to prevent the overfitting at the stage of the sensor selection by setting a positive regularization parameter.

\subsection{OMP, SOMP}
OMP \cite{pati1993orthogonal, tropp2004greed} and simultaneous OMP (SOMP) \cite{tropp2006algorithms} are sparse-approximation methods for single and multiple target variables, respectively.
They can be used for sensor selection for example in \cite{inoba2022optimization}.

The computational complexity of the original algorithm \cite{tropp2006algorithms} of SOMP in the Frobenius-norm criterion is of the forth order as reported in \cite{maung2014improved}.
However, it can be reduced to be of the third order shown in Table~\ref{tb: complexity} by deriving difference equations in a similar way to that in this paper.

SOMP is a greedy method based on linear regression model \eqref{eq: linear_regression_model} with the arbitrary target vector $y$ similarly to GREG.
However, there are two distinct differences.
At each iteration of the greedy selection, SOMP selects an element $i$ that minimizes the approximated estimation error
\begin{align}
  \min_{v \in \mathbb{R}^{N_y}} \norm{E(S) - v X_i}_{\mathrm{F}}^2, \notag
\end{align}
where $E(S) = Y (I_{M} - X_S^\top (X_S X_S^\top)^{-1} X_S)$ is the error matrix for the temporary solution $S$.
On the other hand, GREG selects an element $i$ that minimizes
\begin{align}
  -J(S \cup \{i\}) = \min_{K \in \mathbb{R}^{N_y \times (\abs{S} + 1)}} \norm{Y - K X_{S \cup \{i\}}}_{\mathrm{F}}^2
  + \lambda \norm{K}_{\mathrm{F}}^2. \notag
\end{align}
The first difference is clearly the existence of the regularization.
The second difference is the type of the greedy method: GREG is the direct type while SOMP is not.
GREG selects an element maximizing its objective function $J(S \cup \{i\})$.
On the other hand, SOMP does not necessarily select an element minimizing its objective function $\norm{E(S \cup \{i\})}_{\mathrm{F}}^2$.

\subsection{DG, BDG}
Manohar et~al. proposed a fast sensor-selection algorithm for reconstruction using QR pivoting of the data matrix \cite{manohar2018data}.
This algorithm is reformulated and naturally extended to a greedy algorithm named as determinant-based greedy (DG) sensor selection \cite{saito2021determinant}.

In DG, the following SVD model is assumed:
\begin{align}
    x_S = \Phi_{S, 1:r} w + \tilde{\epsilon},
    \label{eq: SVD_model}
\end{align}
where $r \in \mathbb{N}$ is the truncated number of the SVD, $\Phi_{\cdot, 1:r} \in \mathbb{R}^{N \times r}$ is the left singular matrix of of the $r$ dominant SVD modes, $w \coloneqq \Phi_{\cdot, 1:r}^\top x \in \mathbb{R}^r$ is the vector of the $r$ modes, and $\tilde{\epsilon}$ is the error term.
Vector $w$ is the reduced-order one of the all-measurement vector $x$, and $w$ is estimated from $x_S$.
Based on the linear inverse problem, an optimal estimate of $w$ is given by
\begin{align}
    \hat{w} =  \left\{
    \begin{array}{ll}
      \Phi_{S, 1:r}^\top (\Phi_{S, 1:r} \Phi_{S, 1:r}^\top)^{-1} x_S & p \le r \\
      (\Phi_{S, 1:r}^\top \Phi_{S, 1:r})^{-1} \Phi_{S, 1:r}^\top x_S & p > r
    \end{array}
  \right..
  \label{eq: w_hat}
\end{align}
The objective function of DG measures the accuracy of the estimate $\hat{w}$.
The objective function is
\begin{align}
  J^{\mathrm{DG}}(S) = \left\{
    \begin{array}{ll}
      \det (\Phi_{S, 1:r} \Phi_{S, 1:r}^\top) & p \le r \\
      \det (\Phi_{S, 1:r}^\top \Phi_{S, 1:r}) & p > r
    \end{array}
  \right..
  \label{eq:DG}
\end{align}
The objective function $J^{\mathrm{DG}}$ corresponds to the D-optimality criterion in optimal experimental design \cite{atkinson2007optimum}, in which a set of parameters for linear regression is selected in order to reduce experiment runs.
When $\tilde{\epsilon}$ is white Gaussian noise with zero mean and an uncorrelated covariance matrix, the D-optimality criterion evaluates the inverse volume of the confidence ellipsoid of the estimated latent vector.
The A- and E-optimality criteria also evaluate the size of the confidence ellipsoid with other metrics.
Here, the determinant operator in \eqref{eq:DG} is replaced by the trace-inverse and minimum eigenvalue operators in the A- and E-optimality criteria, respectively.
The greedy sensor-selection algorithms based on the A- and E-optimality criteria, named AG and EG, are proposed in \cite{nakai2021effect}.

Sensor selection by DG is very fast thanks to QR pivoting with the complexity $O(N r^2)$ in the underdetermined case $(p \le r)$.
The computational complexity is increased to be $O(N r^2 p)$ in the overdetermined case $(p > r)$.

Bayesian DG (BDG) \cite{yamada2021fast} is an extension of DG to sensor selection for Bayesian estimation with a correlated noise.
Truncated SVD modes that are not taken into account in the objective function in DG sometimes cause performance deterioration.
In BDG, measurements $x_S$ are assumed to be affected by a noise with a sample covariance matrix of the truncated SVD modes.
In the SVD model \eqref{eq: SVD_model}, it is assumed that $w$ is a Gaussian random vector and $\tilde{\epsilon}$ is a white Gaussian noise with the sample covariance matrix.
An optimal estimate of $w$ different from \eqref{eq: w_hat} is found by posterior estimation under the above assumptions \cite{yamada2021fast}.
The objective function evaluating the inverse volume of the confidence ellipsoid of the optimal estimate in BDG is given by
\begin{align}
  \! J^{\mathrm{BDG}}(S) & = \det \bigl\{
    W W^{\top}
    - W  X_S^{\top} (X_S X_S^{\top})^{-1} X_S W^\top
  \bigr\}, \!
  \label{eq:BDG}
\end{align}
where $W \in \mathbb{R}^{r \times M}$ is the data matrix of the $r$ dominant SVD modes of $X$.

There is a similarity of the objective function between BDG and GREG.
Since the trace is a linear operator and $W$ is constant, the reduced objective function $J_W(S)$ in GREG is equivalent to
\begin{align}
  \tilde{J}_{W}(S) & \coloneqq \trace \bigl( W W^{\top} \bigr) - J_W(S) \notag \\ 
  & = \trace \left\{
    W W^{\top}
    - W  X_S^{\top} (X_S X_S^{\top} + \lambda I_{\abs{S}})^{-1} X_S W^\top
  \right\}.
  \notag 
\end{align}
Hence, the objective function of GREG with $(Y, \lambda) = (W, 0)$ can be regarded as a trace version of that of BDG.

Note that BDG is designed for sensor selection for reconstruction, where the reduced-order matrix of $X$ is estimated, and therefore is not good at sensor selection for general estimation.
BDG has larger order of the computational complexity than the present algorithm as shown in Table~\ref{tb: complexity}.


\section{Applications \label{sec: application}}
We demonstrate the effectiveness of GREG for artificial datasets and a real-world dataset in this section.
Here, GREG is compared to the other data-driven sensor selection algorithms: REG, SOMP, DG, and BDG.

\subsection{Sensor selection for estimation in linear systems}
We apply the data-driven sensor selection algorithms to artificial datasets produced by linear systems.
The linear systems are written as
\begin{align}
    & x_n = C z_n + v_n, \label{eq: LS1} \\
    & y_n = B z_n, \label{eq: LS2}
\end{align}
where $x_n \in \mathbb{R}^{N}$ is the measurements of all candidate sensors, $y_n \in \mathbb{R}^{N_y}$ is the target vector, $z_n  \in \mathbb{R}^{N}$ is the latent variable following an independent and identical distributed Gaussian distribution with mean zero and covariance matrix $\diag (\sigma_{z}^{0}, \sigma_{z}^{2}, \dots, \sigma_{z}^{2 (N - 1)})$ with some positive parameter $\sigma_{z}$ smaller than $1$, and $w_n$ is the white Gaussian noise with mean zero and covariance matrix $\sigma_{v}^2 I_{N}$ with some positive parameter $\sigma_{v}$.
The matrices $C \in \mathbb{R}^{N \times N}$ and $B \in \mathbb{R}^{N_y \times N}$ satisfying $C C^{\top} = I_N$ and $B B^{\top} = I_{N_y}$ are generated by applying QR decomposition to randomly generated matrices.

The latent vector $z_n$ has the shrinking variance $(\sigma_{z}^{0}, \sigma_{z}^{2}, \dots, \sigma_{z}^{2 (N - 1)})$ with the element index increasing for $0 < \sigma_{z} < 1$.
The parameter $\sigma_{z}$ determines the shrinking speed.
For small $\sigma_{z}$, the variance shrinks rapidly, and $z_n$ statistically behaves like a lower-dimension vector because almost variances are nearly zero.
Here, we measure a statistical dimension of $z_n$ as follows:
\begin{align}
    & \hat{r} = \argmin_{m \in \mathbb{N}} \abs{V_{\mathrm{R}}(m) - 0.9999}, \\
    & V_{\mathrm{R}}(m) = \frac{\sum_{i = 1}^m \sigma_{z}^{2 (i - 1)}}{\sum_{i = 1}^N \sigma_{z}^{2 (i - 1)}},
    \label{eq: VR}
\end{align}
where $\hat{r} \in \mathbb{N}$ is the statistical dimension, and $V_{\mathrm{R}}(m) \in \mathbb{R}$ is the ratio of the cumulative variances.
The denominator of \eqref{eq: VR} is the sum of the variances of the all elements of $z_n$, and the numerator is the sum of the variances of the leading-$m$ elements.
The statistical dimension is a positive integer at which $V_{\mathrm{R}}$ takes a value closest to 0.9999 among all positive numbers.

The target vector $y_n$ is implicitly linked to the all-measurement vector $x_n$ via the latent vector $z_n$.
Hence, the target vector $y_n$ can be estimated from some elements of $x_n$.
For any test pair $(x^{\mathrm{new}}, y^{\mathrm{new}})$ following \eqref{eq: LS1}--\eqref{eq: LS2}, the objective in this application is to select elements of $x^{\mathrm{new}}$, i.e., sensors such that $y^{\mathrm{new}}$ can be estimated as accurately as possible.
The matrices $B$, $C$, the latent vector $z^{\mathrm{new}}$, and the noise $v^{\mathrm{new}}$ are unknown in the sensor selection and the estimation, the training data $X \coloneqq [x_1, x_2, \dots, x_M]$ and $Y \coloneqq [y_1, y_2, \dots, y_M]$ are available for selecting sensors $S$, and the training data and the selected-measurement vector $x_S^{\mathrm{new}}$ are available for estimating the target vector $y^{\mathrm{new}}$.

Data-driven sensor-selection algorithms GREG (our proposed method), REG, SOMP, DG, and BDG are independently utilized for the sensor selection.
The dimension $r$ of the reduced-order measurement vector is a design parameter for DG and BDG.
We set the statistical dimension $\hat{r}$ for $r$.
After sensor selection, the target vector $y^{\mathrm{new}}$ is estimated from the selected-measurement vector $x_S^{\mathrm{new}}$ in test data for evaluating effectiveness of each sensor-selection algorithm.
The ridge-regression estimation \eqref{eq: linear_regression_model}, \eqref{eq: Kopt} is used for this purpose, and the estimation error is computed for the selected sensors.
As a performance index of the selected sensors, we use the following normalized estimation error:
\begin{align}
    \tilde{F}(S, \tilde{\lambda}_{\mathrm{est}}) = \frac{F(S, \tilde{\lambda}_{\mathrm{est}})}{F(S_{\mathrm{GREG}}(0), 0)},
    \label{eq: NEE}
\end{align}
where $F(S, \tilde{\lambda}_{\mathrm{est}})$ is the estimation error for the test data when the selected sensors are $S$ and the regularization parameter in the ridge regression is $\tilde{\lambda}_{\mathrm{est}}$, and $S_{\mathrm{GREG}}(\tilde{\lambda}_{\mathrm{sel}})$ is the sensor set selected by GREG with the regularization parameter $\tilde{\lambda}_{\mathrm{sel}}$.
Although setting the same regularization parameters in sensor selection and estimation is appropriate for general applications of GREG, the regularization parameters in sensor selection and estimation are distinguished as $\tilde{\lambda}_{\mathrm{sel}}$ and $\tilde{\lambda}_{\mathrm{est}}$, respectively, for analysis here. 
In \eqref{eq: NEE} the estimation error $F(S, \tilde{\lambda}_{\mathrm{est}})$ of interest is normalized by the estimation error $F(S_{\mathrm{GREG}}(0), 0)$ for GREG without regularization.
Hence, if the normalized estimation error is lower (greater) than one, the estimation error of interest is lower (greater) than the estimation error of GREG without regularization.

Five-fold cross-validation is conducted for computing statics of each index, such as the normalized estimation error.
In the five-fold cross-validation, a sample of $(x_n, y_n)$ are generated first, and the sample is divided into five subsamples of equal sample size.
If a subsample is assigned for test data, remaining four subsamples are assigned for training data.
Sensor selection and evaluation are conducted totally five times for five combinations of data assignment.

Four experiments in each of which sensors are selected with some parameters varying are conducted for comprehensive investigation on performance of the sensor-selection algorithms.
Variable parameters in each experiment is shown in Table~\ref{tb: parameters4LS}.
In each experiment, parameters other than variable ones are fixed to a baseline: $\hat{r} = 10$, $N_y = 10$, $\sigma_v = 10^{-3}$, $M = 5000$, $\tilde{\lambda}_{\mathrm{sel}} = \tilde{\lambda}_{\mathrm{est}} = 0$, where $\tilde{\lambda}_{\mathrm{sel}}$ and $\tilde{\lambda}_{\mathrm{est}}$ are the regularization parameters in sensor selection and estimation, respectively.
Parameter $\sigma_z$ is adjusted such that the cumulative variance ratio $V_{\mathrm{R}}$ takes just $0.9999$ at $\hat{r}$.
The numbers of the candidate sensors and the selected sensors are fixed to $N = 10000$ and $p = 10$.

\begin{table}[b]
  \centering
  \caption{Experiments for sensor selection for linear systems.
  In each experiment, parameters other than variable ones are fixed to the baseline: $\hat{r} = 10$, $N_y = 10$, $\sigma_v = 10^{-3}$, $M = 5000$, $\tilde{\lambda}_{\mathrm{sel}} = \tilde{\lambda}_{\mathrm{est}} = 0$.}
  \label{tb: parameters4LS}
  \begin{tabular}{ll}
    \toprule
    Experiment & Variable  \\
    \midrule
    Experiment~1 & $\hat{r} = 1, 2, 5, 10, 20, 50, 100$ \\
    Experiment~2 & $N_y = 1, 2, 5, 10, 20, 50, 100$ \\
    Experiment~3 & $\sigma_v = 10^{-6}, 10^{-5}, \dots, 10^{0}$ \\
    Experiment~4 & \hspace{-6pt}\begin{tabular}{l}$\tilde{\lambda}_{\mathrm{sel}} = 0, 10^{-12}, 10^{-11}, \dots, 10^{-4}$,
    $\tilde{\lambda}_{\mathrm{est}} = \tilde{\lambda}_{\mathrm{sel}}, 0$, \\
    $M = 10, 20, 50, 100, 200, 500, 1000, 2000, 5000$ \end{tabular}\\
    \bottomrule
  \end{tabular}
\end{table}

\begin{figure*}[tb]
  \centering
  \begin{subfigure}{0.3\textwidth}
      \includegraphics[width=\textwidth]{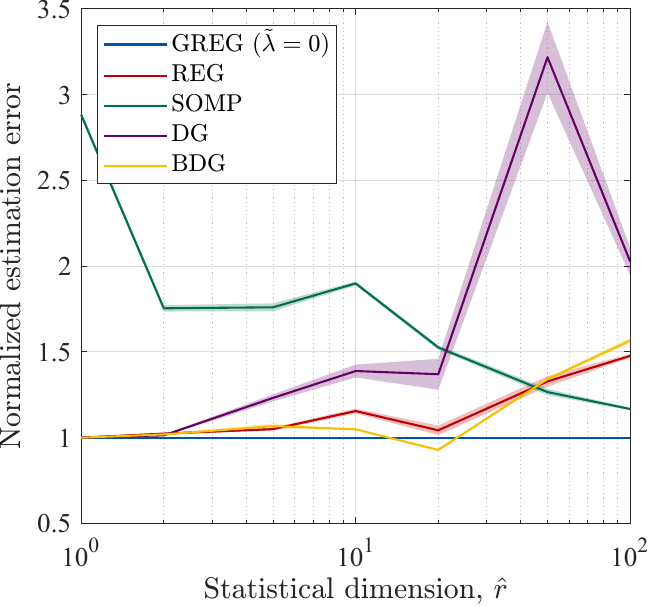}
      \caption{Normalized estimation error and statistical dimension in Experiment~1.}
      \label{fig: LS_error_r}
  \end{subfigure}
  \hfill
  \vspace{5pt}
  \begin{subfigure}{0.3\textwidth}
      \includegraphics[width=\textwidth]{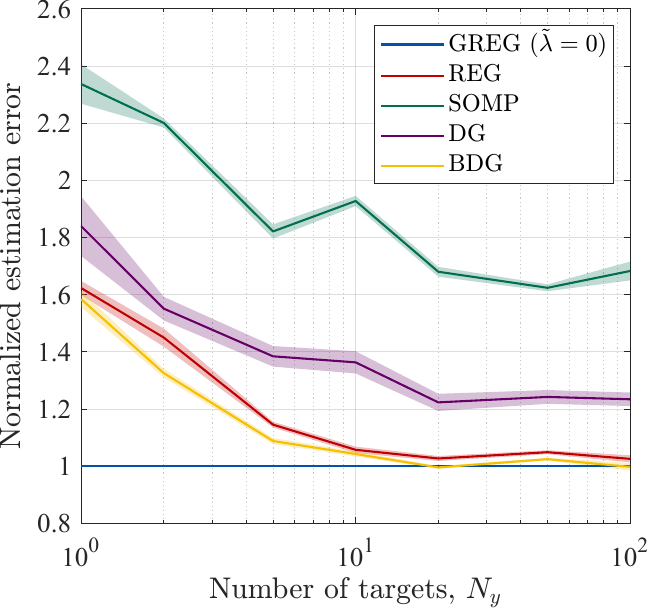}
      \caption{Normalized estimation error and number of targets in Experiment~2.}
      \label{fig: LS_error_Ny}
  \end{subfigure}
  \hfill
  \begin{subfigure}{0.3\textwidth}
      \includegraphics[width=\textwidth]{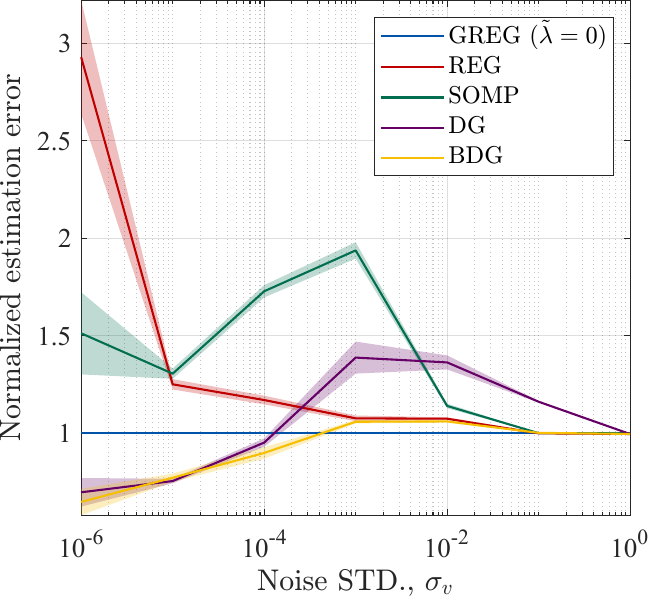}
      \caption{Normalized estimation error and noise standard deviation in Experiment~3.}
      \label{fig: LS_error_sigma}
  \end{subfigure}
  \hfill
  \begin{subfigure}[t]{0.3\textwidth}
      \includegraphics[width=\textwidth]{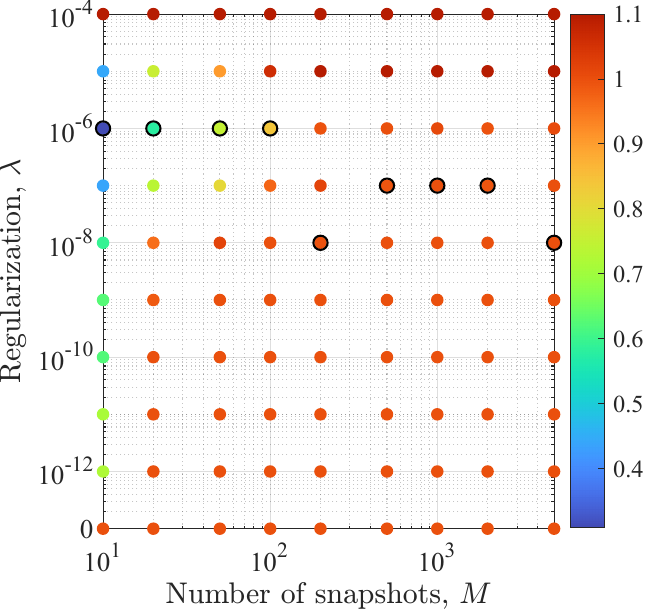}
      \caption{Normalized estimation error of GREG when $\tilde{\lambda}_{\mathrm{sel}} = \tilde{\lambda}_{\mathrm{est}}$ in Experiment~4.}
      \label{fig: LS_error_scatter}
  \end{subfigure}
  \hfill
  \vspace{5pt}
  \begin{subfigure}[t]{0.3\textwidth}
      \includegraphics[width=\textwidth]{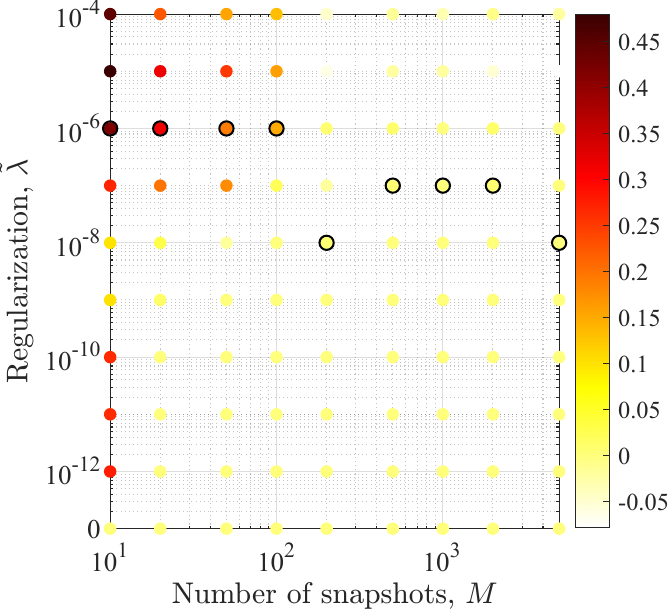}
      \caption{Reduction in normalized estimation error due to considering the regularization parameter in GREG in Experiment~4.}
      \label{fig: LS_contribution_scatter}
  \end{subfigure}
  \hfill
  \begin{subfigure}[t]{0.3\textwidth}
      \includegraphics[width=\textwidth]{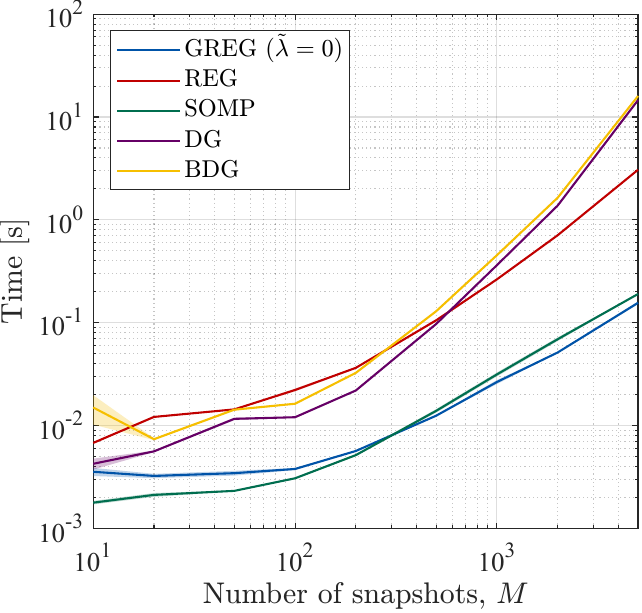}
      \caption{Computation time and number of snapshots in Experiment~4.}
      \label{fig: LS_time}
  \end{subfigure}
  \hfill
  \caption{Results of sensor selection for linear systems.}
  \label{fig: LS}
\end{figure*}

Figure~\ref{fig: LS}(\subref{fig: LS_error_r}) shows relationship between the normalized estimation error~\eqref{eq: NEE} and the statistical dimension $\hat{r}$ in Experiment~1.
The rigid line shows the mean, and the lightly filled area shows the range of the standard error of the mean.
In Fig.~\ref{fig: LS}(\subref{fig: LS_error_r}) GREG without regularization perform best among all sensor-selection algorithms in estimation error except for $\hat{r} = 20$.
The estimation errors of the GREG, REG, and BDG are comparative when the statistical dimension of the latent vector is very small ($\hat{r} = 1, 2$).
The estimation error of GREG is slightly smaller than BDG when $\hat{r} = 5, 10$, and the order of GREG and BDG is reversed when $\hat{r} = 20$.
The estimation error of BDG relatively increases after $\hat{r} = 20$, and the estimation error of SOMP becomes second smallest when $\hat{r} = 50, 100$.

Figure~\ref{fig: LS}(\subref{fig: LS_error_Ny}) shows relationship between the normalized estimation error and the dimension $N_y$ of the target vector.
As $N_y$ decreases, GREG tends to perform better than the reconstruction-oriented algorithms, i.e, REG, DG, and BDG.
GREG performs best when the target vector is low-dimensional $(1 \le N_y \le 10)$.
The estimation errors of REG and BDG approach that of GREG as $N_y$ increases, and the estimation error of BDG is smallest when $N_y = 20$.
The trend of lower estimation error of GREG for relatively small $N_y$ is also observed in Fig.~\ref{fig: LS}(\subref{fig: LS_error_r}), in which $\hat{r}$ is varying with $N_y$ is fixed to $10$.
This trend would be caused because the reconstruction-oriented algorithms selects useless sensors for estimating the target vector.
From \eqref{eq: LS2}, normal modes of the all-measurement vector $x$ to $B$ does not affect the target vector $y$.
Ability to reconstruct such uncorrelated modes does not link to ability to estimate $y$ at all.
When $N_y$ is much smaller than $\hat{r}$, $x$ has higher dimension than the target vector $y$ statistically, and the number of dominant modes of $x$ is larger than $N_y$.
In this case, there is likely to exist many dominant modes of $x$ with low correlation to $y$.
The reconstruction-oriented algorithms can select sensors for reconstructing such low-correlated modes.
These sensors are not effective for estimating $y$.
GREG is designed to select effective sensors for estimating $y$, and therefore GREG performs well even when $N_y$ is small.


Figure~\ref{fig: LS}(\subref{fig: LS_error_sigma}) shows relationship between the normalized estimation error~\eqref{eq: NEE} and the noise standard deviation $\sigma_v$.
On the baseline $\sigma_v = 10^{-3}$, the signal-to-noise ratio of the all-measurement vector is
\begin{align}
    \frac{E[\norm{C z_n}^2]}{E[\norm{v_n}^2]} = \frac{\sigma_z^0 + \dots + \sigma_z^{2 (N - 1)}}{N \sigma_v^2} \simeq 166.
\end{align}
In Fig.~\ref{fig: LS}(\subref{fig: LS_error_sigma}) DG or BDG perform best on the small-noise area, $10^{-6} \le \sigma_v \le 2 \times 10^{-4}$, and GREG performs best on the medium-noise area, $5 \times 10^{-4} \le \sigma_v \le 5 \times 10^{-2}$.
The estimation errors of the sensor selection algorithms except DG are almost equal on the large-noise area, on $10^{-1} \le \sigma_v \le 1$.

Figure~\ref{fig: LS}(\subref{fig: LS_error_scatter}) shows the mean normalized estimation error of GREG when the regularization parameters in selection and estimation are equal $(\tilde{\lambda}_{\mathrm{sel}} = \tilde{\lambda}_{\mathrm{est}} = \tilde{\lambda})$.
The horizontal and vertical axes are the number $M$ of the snapshots and the regularization parameter $\tilde{\lambda}$, respectively.
The black circle shows an optimal regularization parameter which gives the minimum mean estimation error.
The normalized estimation error decreases for the optimal $\tilde{\lambda}$ as $M$ increases, and the estimation error with optimal $\tilde{\lambda}$ is reduced to $31\%$ of that without regularization in particular when $M = 10$.
This would be because the smaller $M$ is, the greater the effect of overfitting, and the greater the effect of regularization.

Figure~\ref{fig: LS}(\subref{fig: LS_error_scatter}) reveals that the estimation error is reduced by setting an appropriate regularization parameter in GREG but does not show how much selection considering the regularization parameter contributes the error reduction.
Even if a sensor set are fixed to one selected by GREG when $\tilde{\lambda}_{\mathrm{sel}} = 0$, $S_{\mathrm{GREG}}(0)$, the estimation error is reduced by setting some positive $\tilde{\lambda}_{\mathrm{est}}$ in estimation.
In order to eliminate the error reduction for a fixed sensor set and evaluate the error reduction only due to sensor selection considering the regularization parameter, we decompose the error reduction as follows:
\begin{align}
    & \tilde{F}(S_{\mathrm{GREG}}(0), 0) - \tilde{F}(S_{\mathrm{GREG}}(\tilde{\lambda}), \tilde{\lambda}) \notag \\
    & \quad = \underbrace{\tilde{F}(S_{\mathrm{GREG}}(0), 0) - \tilde{F}(S_{\mathrm{GREG}}(0), \tilde{\lambda})}_{\mbox{\small reduction due to setting $\tilde{\lambda}$ in estimation}} \notag \\
    & \quad + \underbrace{\tilde{F}(S_{\mathrm{GREG}}(0), \tilde{\lambda}) - \tilde{F}(S_{\mathrm{GREG}}(\tilde{\lambda}), \tilde{\lambda})}_{\mbox{\small reduction due to setting $\tilde{\lambda}$ in GREG}}
    \label{eq: reduction_decomposition}
\end{align}
The first difference of the right hand side of \eqref{eq: reduction_decomposition} is the error reduction due to setting $\tilde{\lambda}_{\mathrm{est}} = \tilde{\lambda}$ in estimation for fixed $S_{\mathrm{GREG}}(0)$.
The second difference is the error reduction due to setting $\tilde{\lambda}_{\mathrm{sel}} = \tilde{\lambda}$ in sensor selection for fixed $\tilde{\lambda}_{\mathrm{est}}$ and our interest.

Figure~\ref{fig: LS}(\subref{fig: LS_contribution_scatter}) shows the reduction in the normalized estimation error due to setting $\tilde{\lambda}$ in GREG, that is, the second difference of \eqref{eq: reduction_decomposition}.
Setting $\tilde{\lambda}$ in GREG results in no change or decrease in the estimation error in many cases.
The error reduction shown in Fig.~\ref{fig: LS}(\subref{fig: LS_contribution_scatter}) takes negative values in 13 cases of $(M, \tilde{\lambda})$, zero in 47 cases, and positive values in 30 cases.
The range of the error reduction is biased in the positive direction, the minimum is $-0.08$, and the maximum is $0.48$. 
The error reduction is $0.42$ for $(M, \tilde{\lambda}) = (10, 10^{-6})$, that is, the estimation error is reduced by $42\%$ due to setting $\tilde{\lambda}$ in GREG in this case.
The positive error reduction is obvious for small $M$ and sensitive to $\tilde{\lambda}$.
This implies that the regularization parameter $\tilde{\lambda}$ in GREG should be optimized when the the sample size $M$ is small.
The optimization of $\tilde{\lambda}$ can be implemented by some numerical methods, such as grid search, which require running GREG multiple times.
Hence, while computation efficiency for large sample size $M$ is of course important, computation efficiency for small $M$ is also important.

Figure~\ref{fig: LS}(\subref{fig: LS_time}) shows the computation time for the sample size $M$ in Test~4.
Here, the results of DG and BDG include the time for SVD, of which time complexity is $O(N M \max \{N, M\})$.
The computation time of GREG increases slowly with increasing $M$, and GREG is fastest when $M \ge 500$.
The computation time of GREG is lower than $4.0 \times 10^{-3}$ s when $M \le 100$, in which overfitting is problematic.
Figure~\ref{fig: LS}(\subref{fig: LS_time}) also reveals that GREG is always faster than REG in Test~4.
This would be because some of $X (\in \mathbb{R}^{10000 \times M})$ in REG are replaced by smaller-size matrix $Y (\in \mathbb{R}^{10 \times M})$ in GREG.
For example, $\norm{X X_i^{\top}}$ for $i = 1, 2, \dots, N$ is computed in the initialization step of REG while $\norm{Y X_i^{\top}}$ with lower complexity is computed in GREG.

\subsection{Sensor selection for yaw-angle estimation of ground vehicle from surface pressure}
We deal with sensor selection for estimating a yaw angle of a ground vehicle from the surface pressure.
This sensor selection was dealt with as a benchmark in \cite{inoba2022optimization, inoba2023improved}.
The yaw angle is important for computing the lateral force on a ground vehicle although it cannot be measured directly in usual running conditions.
The yaw angle is, here, estimated by placing pressure sensors and utilizing pressure at the sensor locations.
The sensor locations are optimized for minimizing the estimation error of the yaw angle by means of the data-driven sensor-selection algorithms in this subsection.
Since a model between the yaw angle and the surface pressure is not obvious beforehand, using GREG, in which the data-driven model is constructed, is reasonable.

The vehicle was an Ahmed model \cite{ahmed1984some} in $1 / 10$ scale and was under freestream velocity of $50$~m/s in a wind tunnel.
Time-average pressure of the upper surface of the vehicle was measured with pressure sensitive paint \cite{liu2021pressure}.
The surface pressure distribution was measured as the yaw angle was changed from $-25^{\circ}$ to $25^{\circ}$ in increments of $1^{\circ}$.
Hence, $51$ pairs of the surface pressure distribution and the yaw angle are available for sensor selection.
See \cite{inoba2023improved} for details of the experimental setup and the measurement data.
Note that the pressure sensitive paint, which was used for acquiring data of the surface pressure distribution, cannot be used outside. 
We consider placing sensors such as piezoresistive ones, at which locations pressure can be measured.

We place sensors symmetrically since the vehicle is bilateral symmetric.
Any pair of the two symmetric sensors assumes to output the difference between the pressures at the two sensor locations.
We hereinafter regard a pair of the symmetric sensors as a single sensor that measures the pressure difference.

\begin{table}[b]
  \centering
  \caption{Settings in sensor selection and estimation for yaw-angle data.}
  \label{tb: parameters4yaw}
  \begin{tabular}{cccccc}
    \toprule
    \multirow{2}{*}{Algorithm} & \multicolumn{3}{c}{Selection} & \multicolumn{2}{c}{Estimation} \\
    \cmidrule(rl){2-4} \cmidrule(rl){5-6}
    & $y$ & $r$ & $\tilde{\lambda}$ & $y$ & $\tilde{\lambda}$  \\
    \midrule
    GREG $(\tilde{\lambda} = 0)$ & $\gamma$ & --    & $0$                & $\gamma$ & $10^{-4}$ \\
    GREG $(\tilde{\lambda} = 10^{-4})$ & $\gamma$ & --    & $10^{-4}$ & $\gamma$ & $10^{-4}$ \\ 
    REG   & $x$      & --    & --                 & $\gamma$ & $10^{-4}$ \\ 
    SOMP  & $\gamma$ & --    & --                 & $\gamma$ & $10^{-4}$ \\ 
    DG    & $w$      & $10$ & --                  & $\gamma$ & $10^{-4}$ \\ 
    BDG   & $w$      & $10$ & --                  & $\gamma$ & $10^{-4}$ \\ 
    \bottomrule
  \end{tabular}
\end{table}

We aim to select sensors so that the yaw angle can be accurately estimated from the measurements of the pressure differences.
The data for the yaw angles $\pm \gamma$ $[^{\circ}]$ are used to evaluate the performance of the sensor-selection algorithms, and the rest data are used to select sensors for each $\gamma \in \{0, 1, \dots, 25\}^{\circ}$.
The data of the pressure distribution and the yaw angles are used, respectively, as $X$ and $Y$ in GREG and SOMP.
Since REG, DG, and BDG are oriented to not estimation but reconstruction, they select sensors not for estimation of the yaw angles but for an accurate reconstruction of the pressure distribution.
The number $p$ of the sensors ranges from $1$ to $10$, and the number of the SVD modes in DG and BDG is $r = 10$.

Since the sample size is small ($M = 49 \mbox{ or } 50$), the overfitting happens easily, and the regularization is expected to be effective.
We apply GREG for each regularization parameter $\tilde{\lambda} = 0, 10^{-10}, 10^{-9}, \dots, 10^{0}$ but show only the results of the two parameters, $\tilde{\lambda} = 0, 10^{-4}$, for ease of visualization.
The results of $\tilde{\lambda} = 0$ are for reference, and value $\tilde{\lambda} = 10^{-4}$ are carefully chosen so that underfitting and overfitting could not cause over $1 \le p \le 10$.
We refer to GREG for $\tilde{\lambda} = 0$ and $\tilde{\lambda} = 10^{-4}$ as GREG $(\tilde{\lambda} = 0)$ and GREG $(\tilde{\lambda} = 10^{-4})$, respectively.
Value $\tilde{\lambda} = 10^{-4}$ is used for the ridge-regression estimation after any sensor selection in order to evaluate the sensor-selection algorithms under the same estimation condition.
The settings in sensor selection and estimation for the present data is summarized in Table~\ref{tb: parameters4yaw}.

\begin{figure*}[tb]
  \centering
  \begin{subfigure}{0.3\textwidth}
      \includegraphics[width=\textwidth]{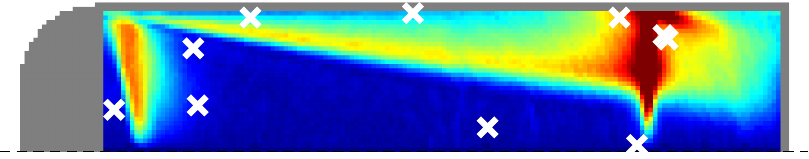}
      \caption{Sensor placement by GREG $(\tilde{\lambda} = 0)$.}
      \label{fig: PSP_EEG1}
  \end{subfigure}
  \vspace{5pt}
  \hfill
  \begin{subfigure}{0.35\textwidth}
      \centering
      \includegraphics[width=.871\textwidth]{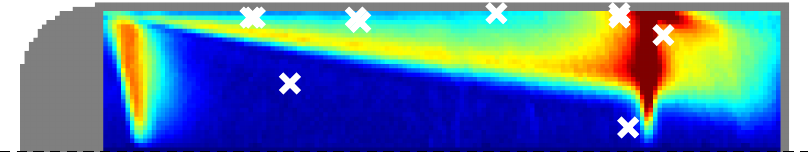}
      \caption{Sensor placement by GREG $(\tilde{\lambda} = 10^{-4})$.}
      \label{fig: PSP_EEG2}
  \end{subfigure}
  \hfill
  \begin{subfigure}{0.3\textwidth}
      \includegraphics[width=\textwidth]{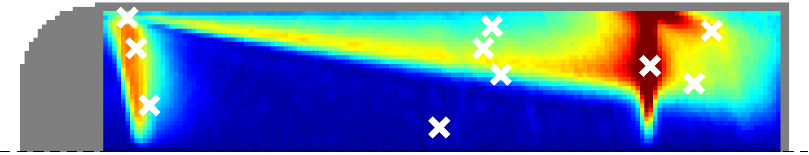}
      \caption{Sensor placement by REG.}
      \label{fig: PSP_REG}
  \end{subfigure}
  \hfill
  \begin{subfigure}{0.3\textwidth}
      \includegraphics[width=\textwidth]{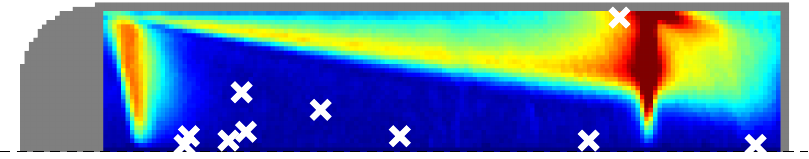}
      \caption{Sensor placement by SOMP.}
      \label{fig: PSP_SOMP}
  \end{subfigure}
  \hfill
  \vspace{5pt}
  \begin{subfigure}{0.3\textwidth}
      \includegraphics[width=\textwidth]{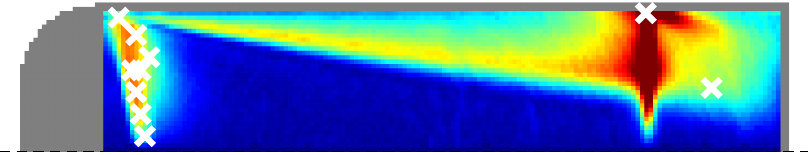}
      \caption{Sensor placement by DG.}
      \label{fig: PSP_DG}
  \end{subfigure}
  \hfill
  \begin{subfigure}{0.3\textwidth}
      \includegraphics[width=\textwidth]{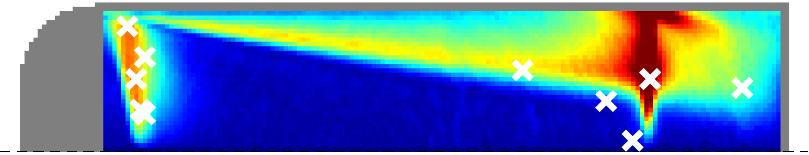}
      \caption{Sensor placement by BDG.}
      \label{fig: PSP_BDG}
  \end{subfigure}
  \hfill
  \begin{subfigure}{0.3\textwidth}
      \includegraphics[width=\textwidth]{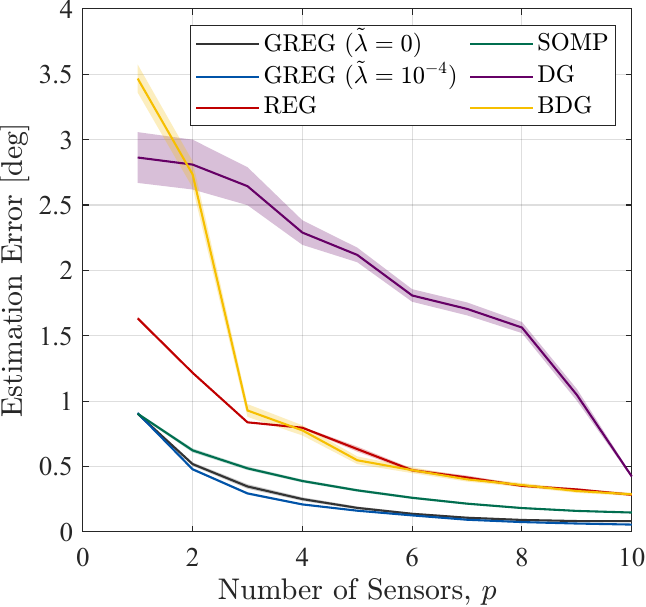}
      \caption{Estimation error for training data.}
      \label{fig: PSP_error_tr}
  \end{subfigure}
  \hfill
  \begin{subfigure}{0.3\textwidth}
      \includegraphics[width=\textwidth]{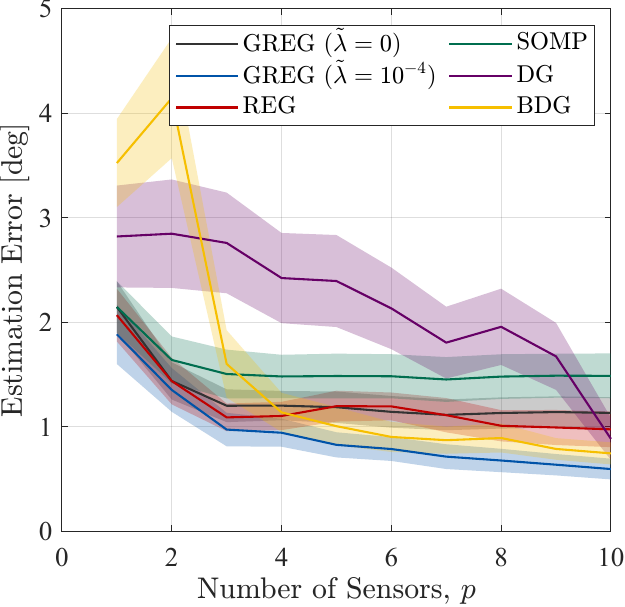}
      \caption{Estimation error for test data.}
      \label{fig: PSP_error}
  \end{subfigure}
  \hfill
  \vspace{5pt}
  \begin{subfigure}{0.3\textwidth}
      \includegraphics[width=\textwidth]{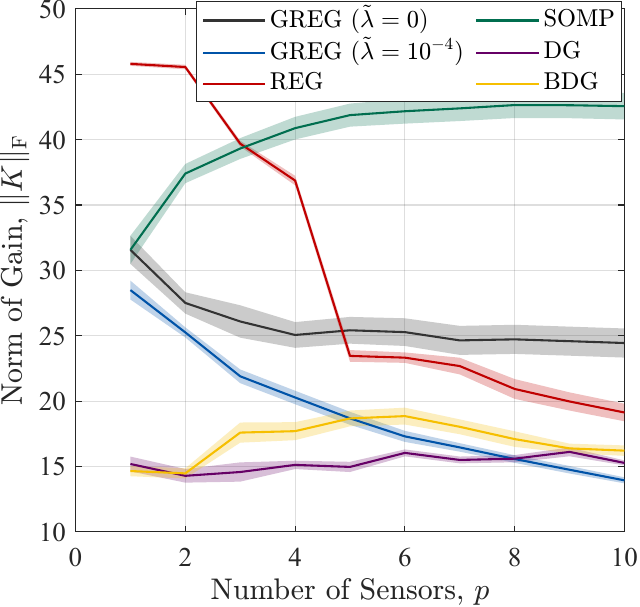}
      \caption{Frobenius norm of estimation gain.}
      \label{fig: PSP_normK}
  \end{subfigure}
  \caption{Results of sensor selection for estimation of vehicle angle.}
  \label{fig: PSP}
\end{figure*}

Figures~\ref{fig: PSP}(\subref{fig: PSP_EEG1})--\ref{fig: PSP}(\subref{fig: PSP_BDG}) show sensors selected by the sensor-selection methods with the white markers when $p = 10$.
The color contour represents the magnitude of the change in the pressure difference between the corresponding location and its symmetric location. The warm and cold colors represent large and small pressure differences, respectively. 
REG, DG, and, BDG, which are tailored to reconstruction of the pressure difference $x$, are likely to select sensors at which the magnitude of the change in $x$ is relatively large.
Such sensor selection would yield measurement with high signal-to-noise ratio.
In contrast, GREG and SOMP, which are for estimating the yaw angle $\gamma$, are not likely to select sensors in the large-magnitude region.
GREG tends to select sensors near the side edge while SOMP selects many sensors around the centerline.
GREG $(\tilde{\lambda} = 10^{-4})$, which takes the regularization into account, selects more pairs of adjacent sensors than GREG $(\tilde{\lambda} = 0)$.
The sensor selection of GREG $(\tilde{\lambda} = 10^{-4})$ sacrifice broad measurements a little but serves robust measurement to the observation noise at the adjacent sensors.

Figure~\ref{fig: PSP}(\subref{fig: PSP_error_tr}) shows the root-mean-square estimation error of the yaw angle for the training data.
The training error is fist, second, and third smallest for GREG $(\tilde{\lambda} = 10^{-4})$, GREG $(\tilde{\lambda} = 0)$, and SOMP, respectively, and the differences among them are small.
The training error for REG and BDG is as small as each other when $p \ge 3$.
DG is worst with respect to the training error.

The relationship that holds in the training error is not always true in the test error shown in fig.~\ref{fig: PSP}(\subref{fig: PSP_error}).
Even though there are some overlaps of the standard-error ranges, GREG $(\tilde{\lambda} = 10^{-4})$ outperforms the other methods also in the test error on average.
The test error for GREG $(\tilde{\lambda} = 0)$ and SOMP is saturated and becomes higher at $p \ge 4$ than that for BDG which becomes the second lowest and keeps decreasing at $p \ge 5$.
Moreover, SOMP has the poorest performance at $p = 10$.
These results suggest low generalization performance of the estimation laws resulting from sensor selection by GREG $(\tilde{\lambda} = 0)$ and SOMP. Meanwhile, it should be noted that the test error of BDG at $p \ge 4$ is unexpectedly low although its objective function for sensor selection is not for estimation of the yaw angle, but for data reconstruction. This was also observed in \cite{inoba2023improved}.

The low generalization performance for GREG $(\tilde{\lambda} = 0)$ and SOMP seems to be due to the large estimation gain due to overfitting. GREG $(\tilde{\lambda} = 0)$ and SOMP result in the larger estimation gain than the other methods when the number of the selected sensors is large as shown in Fig.~\ref{fig: PSP}(\subref{fig: PSP_normK}).
Since the gain magnitude for DG and BDG is low, DG and BDG are unlikely to cause overfitting. 
However, the fitting for the training data is insufficient, and therefore, the fitting for the test data is also poor when the selected sensors are a few. Those characteristics of DG and BDG seem to be because the sensors of DG and BDG are selected not directly for the target yaw angle, but for the reconstruction of data themselves. 
Meanwhile, GREG $(\tilde{\lambda} = 10^{-4})$ attains sensor selection resulting in the good fitting for the training data as well as the small gain in the estimation law.
This contributes to the high generalization performance in GREG $(\tilde{\lambda} = 10^{-4})$.

\section{Conclusion \label{sec: conclusion}}
We proposed a data-driven greedy sensor-selection algorithm, named as GREG, for accurate estimation of the target variables from the selected measurements.
GREG is a natural extension of REG \cite{li2021efficient}, which is a sensor selection algorithm for reconstruction, to one for estimation of the target variables.
GREG greedily selects sensors for minimization of the cost function of the ridge-regression estimator.
The greedy solution is efficiently computed via the recurrence relations, and the time complexity is of the third order.
In sensor selection for artificial linear systems, GREG does not outperform existing data-driven sensor selection algorithms in some conditions in the estimation error, but the higher performance of GREG is obvious in other conditions.
For example, the estimation error of GREG tends to be small when the dimension of the target vector is smaller than the statistical dimension of the latent vector.
Furthermore, it was confirmed that setting a positive regularization not only in linear estimation but also in GREG parameter yielded further reduction in the estimation error when the sample size is small.
In sensor selection for estimation of the yaw angle of the vehicle, GREG with a positive regularization parameter successfully prevented overfitting and outperformed OMP, which is a selection algorithm for the linear regression without regularization, for the test data.

The regularization parameter in GREG should be optimized if overfitting is problematic.
The optimization generally requires running GREG multiple times.
Fortunately, when overfitting is problematic in linear estimation, the sample size tends to be small, and running GREG would not be time-consuming.
A practical issue, here, is how we know whether the optimization in GREG is needed or not.
A heuristic solution to this is running GREG with theregularization parameter of zero, and then, optimizing the regularization parameter in ridge regression for the selected sensors.
If the test error is greatly reduced by the optimization in ridge regression, then the optimization of the regularization parameter in GREG would be also effective.
The optimization in ridge regression is generally much faster than that in GREG, and thus the former should be tried first.

Target quantities are often estimated from other measured quantities in engineering and scientific fields. For example, flow velocity distributions are estimated from sound measured by a microphone in the fluid engineering field \cite{ozawa2022spatiotemporal}, and emotion states are estimated from brain signals measured by electrodes on a head in the brain/neuro science field \cite{miyamoto2021emotion}. For these estimation tasks, GREG expects to provide effective placements of microphones and electrodes. GREG also seems to be useful for sensor selection for prediction, such as tidal current prediction and air pollution prediction. For this, the target vector is set to be a time series of future tidal velocities or concentrations of pollution particles.

Devices equipped with several sensors and particle image velocimetry in fluid engineering measure multiple physical quantities per location.
GREG is not suitable for optimizing measurement locations of such devices and measurement methods because single physical quantity per location is assumed to be measured in design of GREG.
Extending GREG to an optimization algorithm that selects locations of the above measurement systems will be the future work. 


\appendices
\label{sec: appendix}

\section{Derivations of some relations}
\subsection{Lemma of difference equations of posterior-covariance-like matrices}
\label{sec: lm_incQ}
Difference equations of $Q^{xx}$, $Q^{xy}$, and $Q^{yy}$, which are shown in the following lemma, will play an important role on derivation of the fast greedy algorithm.
\begin{lemma}
  \label{lm: incQ}
  For any $S \subset \Omega$ and $i \in \Omega \setminus S$ so that $P^{xx}_{S \cup \{i\}, S \cup \{i\}} \succ 0$, the following equations hold:
  \begin{align}
    &
    Q^{x x}(S) - Q^{x x}(S \cup \{ i \})
    = \notag \\
    & \hspace{80pt} (Q^{x x}_{i} (S))^{\top} (Q^{x x}_{i, i} (S))^{-1} Q^{x x}_{i} (S)
    \label{eq: incQxx} \\
    &
    Q^{x y}(S) - Q^{x y}(S \cup \{ i \})
    = \notag \\
    & \hspace{80pt} (Q^{x x}_{i} (S))^{\top} (Q^{x x}_{i, i} (S))^{-1} Q^{x y}_{i} (S)
    \label{eq: incQxy} \\
    &
    Q^{y y}(S) - Q^{y y}(S \cup \{ i \})
    = \notag \\
    & \hspace{80pt} (Q^{x y}_{i} (S))^{\top} (Q^{x x}_{i, i} (S))^{-1} Q^{x y}_{i} (S).
    \label{eq: incQyy}
  \end{align}
\end{lemma}
\begin{proof}
  We show that (\ref{eq: incQyy}) holds.
  Matrix $P^{xx}_{S \cup \{ i \}, S \cup \{ i \}}$ can be inverted blockwise as follows:
  \begin{align}
    (P^{x x}_{S \cup \{ i \}, S \cup \{ i \}})^{-1} =
    &
    \left[
      \begin{IEEEeqnarraybox*}[][c]{,c/c,}
        P^{x x}_{S, S} & P^{x x}_{S, i}\\
        P^{x x}_{i, S} & P^{x x}_{i, i}
      \end{IEEEeqnarraybox*}
    \right]^{-1}
    \notag \\ =
    &
    \left[
      \begin{IEEEeqnarraybox*}[][c]{,c/c,}
        (P^{x x}_{S, S})^{-1} & 0 \\
        0 & 0
      \end{IEEEeqnarraybox*}
    \right]
    +
    \left[
      \begin{IEEEeqnarraybox*}[][c]{,c,}
        - (P^{x x}_{S, S})^{-1} P^{x x}_{S, i} \\
        1 
      \end{IEEEeqnarraybox*}
    \right]
    \notag \\
    &
    (Q^{x x}_{i, i}(S))^{-1}
    \left[
      \begin{IEEEeqnarraybox*}[][c]{,c,}
        - (P^{x x}_{S, S})^{-1} P^{x x}_{S, i} \\
        1
      \end{IEEEeqnarraybox*}
    \right]^{\top}
    \label{eq: tmp_incQyy}
  \end{align}
  Note that $Q^{x x}_{i, i}(S)$ is the Schur complement of $P^{x x}_{i, i}$ in $P^{x x}_{S \cup \{ i \}, S \cup \{ i \}}$.
  Multiplying (\ref{eq: tmp_incQyy}) by $-(P^{x y}_{S \cup \{ i \}})^{\top}$ and $P^{x y}_{S \cup \{ i \}} $ from the left and the right, respectively, and adding $P^{yy}$ yield
  \begin{align}
    Q^{y y}(S \cup \{ i \})
    = Q^{y y}(S) - (Q^{x y}_{i} (S))^{\top} (Q^{x x}_{i, i} (S))^{-1} Q^{x y}_{i} (S). \notag
  \end{align}
  Equations (\ref{eq: incQxx}) and (\ref{eq: incQxy}) can be derived in a similar way.
\end{proof}
Equations~\eqref{eq: incQxx}--\eqref{eq: incQyy} are the difference equations of $Q_{xx}$, $Q_{xy}$, and $Q_{yy}$.
If $Q_{xx}(S)$, $Q_{xy}(S)$, and $Q_{yy}(S)$ are known, $Q_{xx}(S \cup \{ i \})$, $Q_{xy}(S \cup \{ i \})$, and $Q_{yy}(S \cup \{ i \})$ can be computed via \eqref{eq: incQxx}--\eqref{eq: incQyy}.
Updating $Q_{xx}$, $Q_{xy}$, and $Q_{yy}$ via \eqref{eq: incQxx}--\eqref{eq: incQyy} is much faster than direct computation of $Q_{xx}(S \cup \{ i \})$, $Q_{xy}(S \cup \{ i \})$, and $Q_{yy}(S \cup \{ i \})$ because the matrix products and inversion are not required in \eqref{eq: incQxx}--\eqref{eq: incQyy}.
For example, the decrease in $Q^{yy}$ can be evaluated by multiplications of vectors $Q^{x y}_{i} (S)$ and inverse of scalar $Q^{x x}_{i, i} (S)$ according to \eqref{eq: incQyy}.

\subsection{Proof of Theorem~\ref{thm: incJ}}
\label{sec: pf_incJ}
The objective function $J(S)$ can be written with the use of the covariance-like matrices as
\begin{align}
  J(S) = \trace \bigl\{ (P^{x y}_S)^{\top} (P^{x x}_{S, S})^{-1} P^{x y}_S \bigr\}
   = \trace \bigl\{ P^{y y} - Q^{y y}(S) \bigr\}. \notag
\end{align}
Therefore, the difference of $J$, which appears in Algorithm~\ref{alg: greedy}, is written by
\begin{align}
  J(S \cup \{ i \}) - J(S) = \trace \bigl\{ Q^{y y}(S) - Q^{y y}(S \cup \{ i \}) \bigr\}.
  \label{eq: incJ_pre}
\end{align}
This equation means that increase in $J$ is the trace of the decrease in $Q^{yy}$.

\subsection{Proof of Lemma~\ref{lm: xi_theta}}
\label{sec: xi_theta}
We prove this lemma by mathematical induction.
When $k = 1$, from definitions (\ref{eq: delta})--(\ref{eq: theta}), we have
\begin{align}
  \delta^{(1)} & = X X_{\bar{s}_1}^{\top} + \lambda \, (I_{N})_{\bar{s}_1}^{\top}
  = {Q^{xx}_{\bar{s}_1}(\bar{S}_0)}^{\top}
  \notag \\
  \xi^{(1)} & = \delta^{(1)} / \sqrt{\delta^{(1)}_{\bar{s}_1}}
  = {Q^{xx}_{\bar{s}_1}(\bar{S}_0)}^{\top} ({Q_{\bar{s}_1, \bar{s}_1}^{xx}(\bar{S}_{0})})^{-1/2}
  \notag \\
  \theta^{(1)} & = Y X_{\bar{s}_1}^{\top} / \sqrt{\delta^{(1)}_{\bar{s}_1}}
  = {Q^{xy}_{\bar{s}_1}(\bar{S}_0)}^{\top} ({Q_{\bar{s}_1, \bar{s}_1}^{xx}(\bar{S}_{0})})^{-1/2}
  \notag
\end{align}
By setting $S = \bar{S}_0$ and $i = \bar{s}_1$ in (\ref{eq: incQxx}) and (\ref{eq: incQxy}), we get
\begin{align}
  Q^{xx}(S_1) & = Q^{xx}(\bar{S}_0) - (Q^{x x}_{\bar{s}_1} (\bar{S}_0))^{\top} (Q^{x x}_{\bar{s}_1, \bar{s}_1} (\bar{S}_0))^{-1} Q^{x x}_{\bar{s}_1} (\bar{S}_0) \notag \\
  & = Q^{xx}(\bar{S}_0) - \xi^{(1)} {\xi^{(1)}}^{\top},
  \label{eq: incQxx2k1}\\
  Q^{x y}(S_1) & = Q^{x y}(\bar{S}_0) - (Q^{x x}_{\bar{s}_1} (\bar{S}_0))^{\top} (Q^{x x}_{\bar{s}_1, \bar{s}_1} (\bar{S}_0))^{-1} Q^{x y}_{\bar{s}_1} (\bar{S}_0) \notag \\
  & = Q^{xy}(\bar{S}_0) - \xi^{(1)} {\theta^{(1)}}^{\top}.
  \label{eq: incQxy2k1}
\end{align}
Equations (\ref{eq: incQxx3}) and (\ref{eq: incQxy3}) for $k = 1$ is obtained by using definitions $Q^{xx}(\bar{S}_0) = X X^{\top} + \lambda \, I_{N}$ and $Q^{xy}(\bar{S}_0) = X Y^{\top}$ in (\ref{eq: incQxx2k1}) and (\ref{eq: incQxy2k1}), respectively.
Now, we show that, if (\ref{eq: omega2})--(\ref{eq: incQxy3}) hold for any $k = l$, (\ref{eq: omega2})--(\ref{eq: incQxy3}) hold also for $k = l + 1$.
From (\ref{eq: incQxx3}) and (\ref{eq: incQxy3}) with $k = l$, the following equations hold:
\begin{align}
  & \delta^{(l+1)} = X X_{s_{l+1}}^{\top} + \lambda \, (I_{N})_{s_{l+1}}^{\top} - \Xi^{(l)} (\Xi^{(l)}_{s_{l+1}})^{\top}
  \notag \\
  & \hspace{.86cm} = {Q^{xx}_{s_{l+1}}(S_{l})}^{\top}
  \notag \\
  & Y X_{s_{l+1}}^{\top} - \Theta^{(l)} ({\Xi^{(l)}_{s_{l+1}}})^{\top}
  = {Q^{xy}_{s_{l+1}}(S_{l})}^{\top}
  \notag
\end{align}
By substituting the above equations into (\ref{eq: omega}) and (\ref{eq: theta}), respectively, we get (\ref{eq: omega2}) and (\ref{eq: theta2}) for $k = l + 1$.
From (\ref{eq: omega2}) and (\ref{eq: theta2}), (\ref{eq: incQxx2}) and (\ref{eq: incQxy2}) with $k = l + 1$ are obtained in the same way as the derivation of (\ref{eq: incQxx2k1}) and (\ref{eq: incQxy2k1}), respectively.
Equations (\ref{eq: incQxx3}) and (\ref{eq: incQxy3}) with $k = l + 1$ are derived as follows:
\begin{align}
  Q^{x x}(S_{l+1}) & = Q^{x x}(S_{l}) - \xi^{(l+1)} {\xi^{(l+1)}}^{\top}
  \notag \\
  & = X X^{\top} + \lambda I_{N} - \Xi^{(l)} {\Xi^{(l)}}^{\top} - \xi^{(l+1)} {\xi^{(l+1)}}^{\top}
  \notag \\
  & = X X^{\top} + \lambda I_{N} - \Xi^{(l+1)} {\Xi^{(l+1)}}^{\top},
  \notag \\
  Q^{x y}(S_{l+1}) & = Q^{x y}(S_{l}) - \xi^{(l+1)} {\theta^{(l+1)}}^{\top}
  \notag \\
  & = X Y^{\top} - \Xi^{(l)} {\Theta^{(l)}}^{\top} - \xi^{(l+1)} {\theta^{(l+1)}}^{\top}
  \notag \\
  & = X Y^{\top} - \Xi^{(l+1)} {\Theta^{(l+1)}}^{\top}.
  \notag
\end{align}

\subsection{Proof of Theorem~\ref{thm: update_fg}}
\label{sec: update_fg}
By using (\ref{eq: incQxy2}) and (\ref{eq: incQxy3}), we have
\begin{align}
f_i^{(k)} 
& = \Norm{(Q_i^{x y}(S_{k-1}))^{\top} - \xi_i^{(k)} {\theta^{(k)}}}^2
\notag \\
& = \Norm{(Q_i^{x y}(S_{k-1}))^{\top}}^2 - \xi_i^{(k)} \bigr( 2 Q_i^{x y}(S_{k-1}) \theta^{(k)}
\notag \\
& \hspace{4cm} - \xi_i^{(k)} \Norm{\theta^{(k)}}^2 \bigl)
\notag \\
& = f_i^{(k-1)} - \xi_i^{(k)} \Bigl\{ 2 (X_i Y^{\top} - \Xi_i^{(k-1)} {\Theta^{(k-1)}}^{\top}) \theta^{(k)}
\notag \\
& \hspace{4cm} - \xi_i^{(k)} \Norm{\theta^{(k)}}^2 \Bigr\} \notag
\end{align}
From (\ref{eq: incQxx2}) for $(i, i)$ entry, we get (\ref{eq: incg}).

\subsection{Identification of Feasible Set}
\label{sec: identification}
We give an efficient method to compute the feasible set $\mathcal{I}_k$.
The feasible element $i$, which satisfies $P_{\bar{S}_{k-1} \cup \{i\}, \bar{S}_{k-1} \cup \{i\}}^{x x} \succ 0$, can be found by computing $g_i^{(k-1)}$ according to the following lemma.
\begin{lemma}\label{lm: feasible_element}
  $P_{\bar{S}_{k-1} \cup \{i\}, \bar{S}_{k-1} \cup \{i\}}^{x x} \succ 0$ if and only if $g_i^{(k-1)} > 0$ for any $k \in 1:p$ and any $i \in \Omega \setminus \bar{S}_{k-1}$.
\end{lemma}
\begin{proof}
  The greedy solution $\bar{S}_{k-1}$ satisfies $P_{\bar{S}_{k-1}, \bar{S}_{k-1}}^{x x} \succ 0$. Therefore, it is found that $P_{\bar{S}_{k-1} \cup \{i\}, \bar{S}_{k-1} \cup \{i\}}^{x x} \succ 0$ is equivalent to $g_i^{(k-1)} = P_{i, i}^{x x} - P_{i, \bar{S}_{k-1}}^{x x} P_{\bar{S}_{k-1}, \bar{S}_{k-1}}^{-1} P_{\bar{S}_{k-1}, i}^{x x} > 0$ by applying the Schur complement lemma to $P_{\bar{S}_{k-1} \cup \{i\}, \bar{S}_{k-1} \cup \{i\}}^{x x}$.
\end{proof}
The above lemma shows that the feasible set $\mathcal{I}_k$ is a set of indices $i \in \Omega \setminus \bar{S}_{k-1}$ for each of which $g_i^{(k-1)} > 0$. 
Since, if $P^{xx}_{S, S}$ is not positive definite, $P^{xx}_{S \cup \{i\}, S \cup \{i\}}$ is not positive definite either, the feasible set $\mathcal{I}_k$ is monotone decreasing, that is, $\mathcal{I}_{k-1} \supset \mathcal{I}_{k}$ with $\mathcal{I}_0 \coloneqq \Omega$ for all $k \in 1:p$.
From this monotonicity of $\mathcal{I}_k$ and Lemma~\ref{lm: feasible_element}, the feasible set defined in (\ref{eq: feasible_set}) can be expressed by
\begin{align}
  \mathcal{I}_k = \Set{i \in \mathcal{I}_{k-1} \setminus \bar{S}_{k-1}}{g_i^{(k-1)} > 0}. \notag
\end{align} 
The feasible set $\mathcal{I}_k$ can be found by computing $g_i^{(k-1)}$ for all $i \in \mathcal{I}_{k-1} \setminus \bar{S}_{k-1}$ and collecting elements that satisfy $g_i^{(k-1)} > 0$.

\bibliographystyle{IEEEtran}
\bibliography{xaerolab}

\vspace{-33pt}
\begin{IEEEbiography}
[{\includegraphics[width=1in,height=1.25in,clip,keepaspectratio]{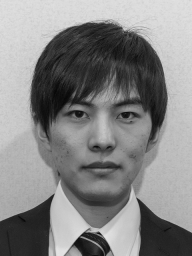}}]{Yasuo Sasaki} received a B.S. degree in mechanical and aerospace engineering from Nagoya University, Japan, in 2017. 
He received M.S. and Ph.D. degrees in aerospace engineering from Nagoya University, Japan, in 2019, and 2022, respectively. 
He is currently an Assistant Professor of the Department of Aerospace Engineering at Nagoya University, Japan. 
\end{IEEEbiography}

\begin{IEEEbiography}
[{\includegraphics[width=1in,height=1.25in,clip,keepaspectratio]{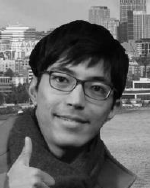}}]{Keigo Yamada} received a B.S. degree in physics from Tohoku University, Japan, in 2019. 
He received M.S. and Ph.D. degrees in aerospace engineering from Tohoku University, Japan, in 2021 and 2024, respectively.
\end{IEEEbiography}

\begin{IEEEbiography}
[{\includegraphics[width=1in,height=1.25in,clip,keepaspectratio]{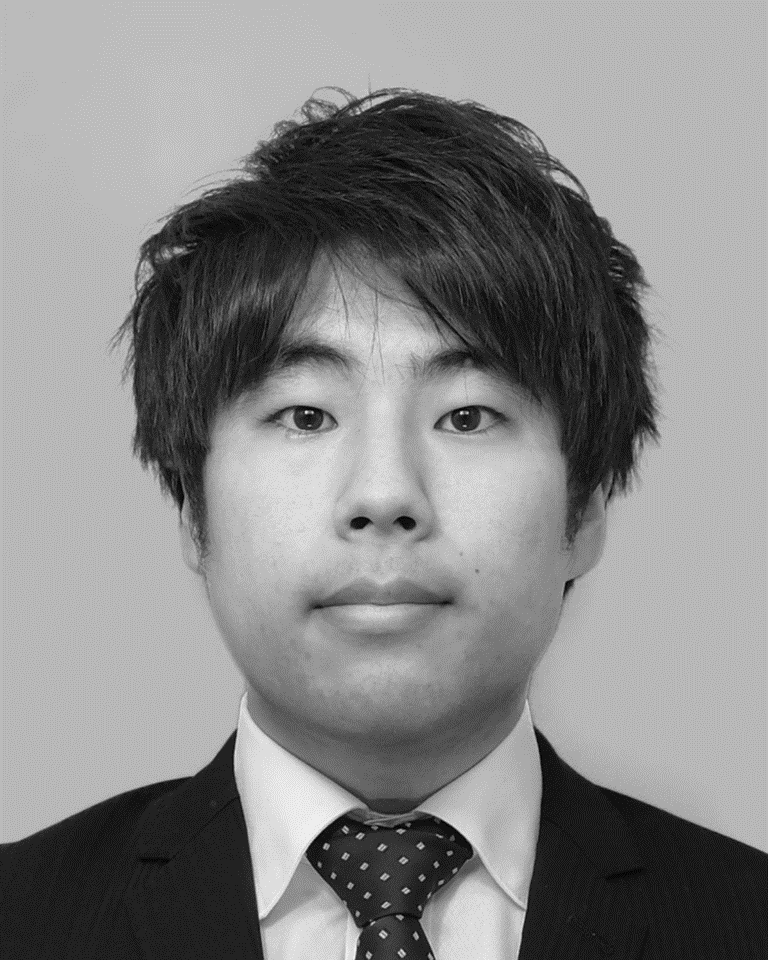}}]{Takayuki Nagata} received B.S. and M.S. degrees in mechanical and aerospace engineering from Tokai University, Japan, in 2015 and 2017, respectively. He received a Ph.D. degree in aerospace engineering from Tohoku University, Japan, in 2020. From 2018 to 2020, he was a Research Fellow of the Japan Society for the Promotion of Science (JSPS) at Tohoku University, Japan. He is currently an Assistant Professor of the Department of Aerospace Engineering at Nagoya University, Japan. 
\end{IEEEbiography}

\begin{IEEEbiography}
[{\includegraphics[width=1in,height=1.25in,clip,keepaspectratio]{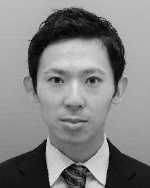}}]{Yuji Saito} received a B.S. degree in mechanical engineering and a Ph.D. degree in mechanical space engineering from Hokkaido University, Japan, in 2014 and 2018, respectively.
He is currently an Assistant Professor with the Frontier Research Institute for Interdisciplinary Sciences, Tohoku University, Japan.
\end{IEEEbiography}

\begin{IEEEbiography}
[{\includegraphics[width=1in,height=1.25in,clip,keepaspectratio]{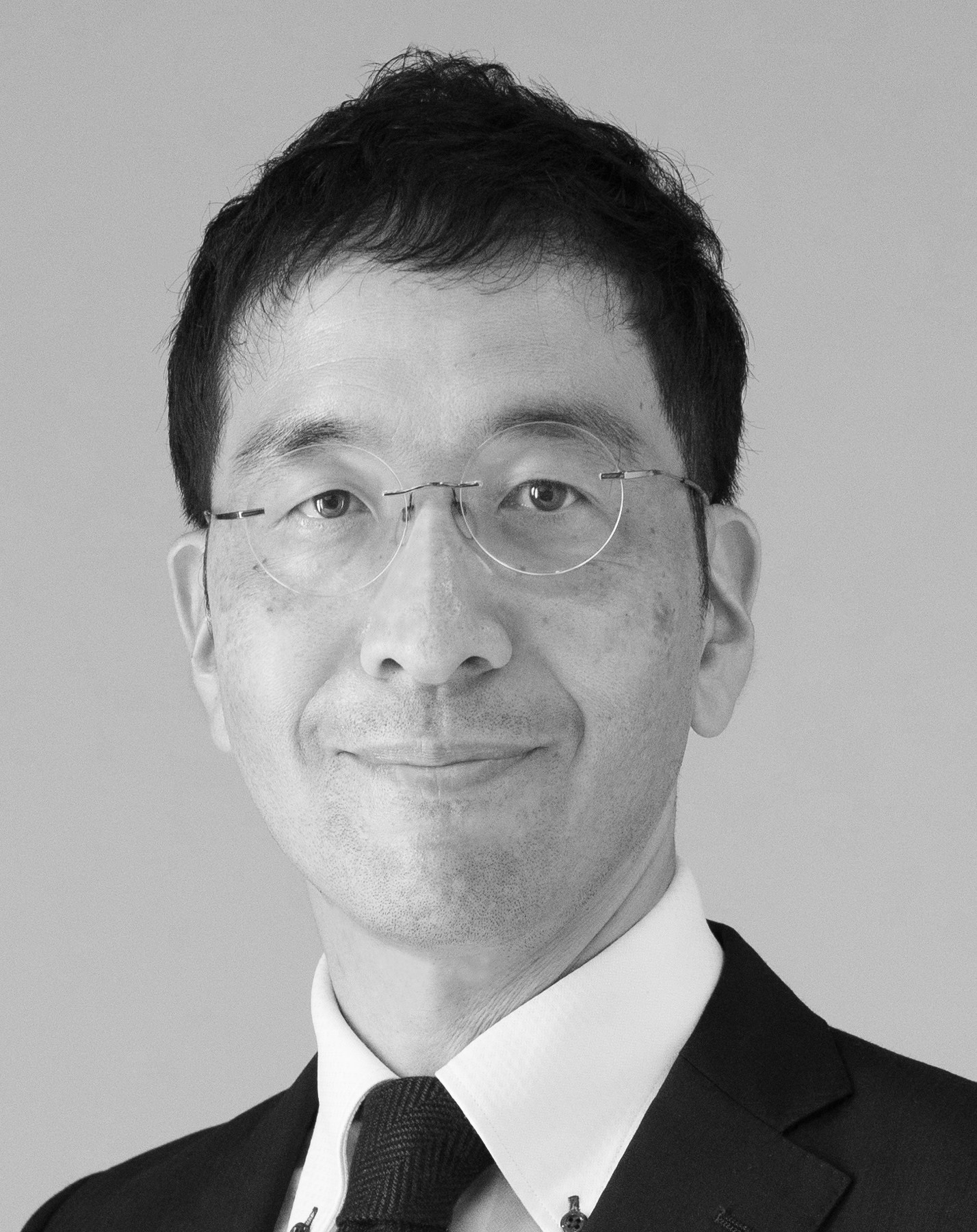}}]{Taku Nonomura} received a B.S. degree in mechanical and aerospace engineering from Nagoya University, Japan, in 2003, and a Ph.D. degree in aerospace engineering from the University of Tokyo, Japan in 2008. He is currently a Professor of the Department of Aerospace Engineering at Nagoya University, Japan.
\end{IEEEbiography}

\vfill

\end{document}